# Emergent Facilitation and Glassy Dynamics in Supercooled Liquids

Muhammad R. Hasyim[a,1] and Kranthi K. Mandadapu[a,b,1]

[a]Department of Chemical and Biomolecular Engineering, University of California, Berkeley, CA 94720; [b]Chemical Sciences Division, Lawrence Berkeley National Laboratory, Berkeley, CA 94720

**In supercooled liquids, dynamical facilitation refers to a phenomenon where microscopic motion begets further motion nearby, resulting in spatially heterogeneous dynamics. This is central to the glassy relaxation dynamics of such liquids, which show super-Arrhenius growth of relaxation timescales with decreasing temperature. Despite the importance of dynamical facilitation, there is no theoretical understanding of how facilitation emerges and impacts relaxation dynamics. Here, we present a theory that explains the microscopic origins of dynamical facilitation. We show that dynamics proceeds by localized bond-exchange events, also known as excitations, resulting in the accumulation of elastic stresses with which new excitations can interact. At low temperatures, these elastic interactions dominate and facilitate the creation of new excitations near prior excitations. Using the theory of linear elasticity and Markov processes, we simulate a model, which reproduces multiple aspects of glassy dynamics observed in experiments and molecular simulations, including the stretched exponential decay of relaxation functions, the super-Arrhenius behavior of relaxation timescales as well as their two-dimensional (2D) finite-size effects. The model also predicts the subdiffusive behavior of the mean squared displacement (MSD) on short, intermediate timescales. Furthermore, we derive the phonon contributions to diffusion and relaxation, which when combined with the excitation contributions produce the two-step relaxation processes, and the ballistic-subdiffusive-diffusive crossover MSD behaviors commonly found in supercooled liquids.**

Supercooled liquids | Glassy dynamics | Dynamical facilitation | Elasticity theory

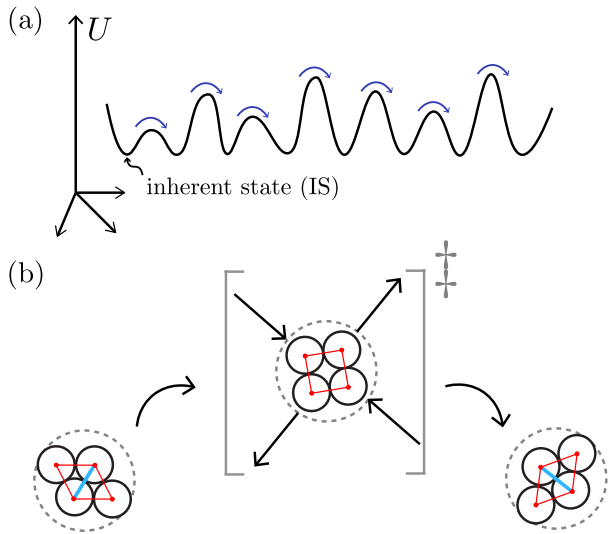

**Fig. 1.** (a) Glassy dynamics as a hopping process in a multi-dimensional potential energy landscape (PEL) whose minima are denoted as the inherent states. (b) Any transition between two nearest neighbor inherent states in the PEL occurs via bond-exchange events, or excitations, where particles break an existing bond and form a new one (cyan). Shown in the middle is a model transition state with arrows indicating the displacements of the particles involved in the bond-exchange event. These bond-exchange events are T1 transitions in 2D systems.

**G**lassy dynamics refers to the dramatic slowdown of microscopic motion in supercooled liquids below an onset temperature $T_o$ (1). The slowdown is accompanied by dynamical heterogeneity (2), a phenomenon where microscopic motion is clustered into regions of high and low mobility (3). In contrast to the slow and heterogeneous dynamics, the structure of supercooled liquids is homogeneous in space and varies little with temperature (4, 5). Such a contrast between structure and dynamics differs from how we understand relaxation dynamics in normal liquids, where structural information, e.g., the radial distribution function (RDF) $g(r)$, plays a foundational role (6). Despite the ubiquity of supercooled liquids, understanding the microscopic mechanisms behind their glassy dynamics remains incomplete.

Over the years, molecular dynamics (MD) simulations have revealed key microscopic observations relevant to understanding how dynamics proceed spatially in supercooled liquids. At short timescales, microscopic motion progresses by the emergence of sparse and spatially localized mobile regions, herein referred to as excitations (3). These excitations populate the system at an average rate $c_\sigma$ that is Arrhenius below $T_o$, i.e., $c_\sigma \sim e^{-\beta J_\sigma}$ where $J_\sigma$ is the activation energy (3, 5). This is followed by the emergence of dynamical facilitation, where regions of high mobility spread over time due to motion begetting nearby motion (3, 7–9). The notion of dynamical facilitation is more apparent in a recent MD study (10), where algorithmic advances allow equilibration of supercooled liquids and their long-time simulations at ultralow temperatures. These observations inspire dynamical facilitation (DF) theory (11, 12) to use the idea of facilitating excitations, implemented in kinetically constrained models (13), to describe different facets of glassy dynamics. But despite their well-known observations and use in DF theory, the microscopic origins behind excitations and facilitation as emergent phenomena have yet to be discovered.

Recent works (5, 14) have delved into the origin of excitations and the onset temperature for glassy dynamics by the linear elasticity theory (15) and the theory of defect-mediated phase transitions (16–20). The work in Ref. (5) demonstrates that excitations are localized motions connecting nearest neighboring inherent states, which are energy-minimizing configurations of the supercooled liquid. These excitations are further shown to be localized pure-shear deformations driven by elementary bond-exchange events, e.g., T1 transitions in

---

two-dimensional (2D) particle systems (Fig. 1). Given the pure-shear nature of the strain fields, linear elasticity theory (15) allows us to compute the energy barriers for motion $J_\sigma$ using the RDF $g(r)$ and the inherent-state shear modulus. Building from Ref. (5), the work in Ref. (14) constructs a theory of onset temperature $T_o$ for glassy dynamics in 2D supercooled liquids, where excitations are viewed as defects in an amorphous medium (21). This defect picture gives rise to a melting scenario for the onset of glassy dynamics, where supercooled liquids lose their inherent rigidity at temperatures $T > T_o$. In 2D, the melting transition fits the Kosterlitz-Thouless-Halperin-Nelson-Young (KTHNY) scenario (16–20), providing a connection between the physics of 2D glassy and crystalline materials.

Even though there exists now the notion of excitations, what remains an open question is the microscopic basis behind the emergence of dynamical facilitation and its impact on the overall relaxation of supercooled liquids. In this work, we use ideas from linear elasticity theory to understand how and why excitations facilitate their own spread in supercooled liquids. To this end, we put forward a theory to show that facilitation emerges from the elastic interaction of an excitation with the stresses accumulated by a history of excitations. By implementing a model for relaxation based on elastically interacting excitations, the theory accounts for the following behaviors of supercooled liquids that any theory of glassy dynamics must explain:

- The stretched exponential decay of relaxation functions, such as the self-intermediate scattering function $F_s(\mathbf{k}, t)$ for a wavevector $\mathbf{k}$, and the bond-order autocorrelation function $C_b(t)$.
- The crossover from high-temperature Arrhenius to super-Arrhenius growth of equilibrium relaxation timescales $\tau_{eq}$ below the onset temperature $T_o$.
- The subdiffusive behavior of the mean-squared displacement (MSD) preceding the long-time diffusion limit.
- In two dimensions (2D), the finite-size effects in relaxation including the contributions from the Mermin-Wagner phonon fluctuations.
- The role of phonon fluctuations in mediating structural relaxation and diffusion, distinguished from the excitations.

Together with Ref. (5), this work presents a self-consistent microscopic picture of glassy dynamics based on emergent facilitation, specializing in two dimensions (2D).

## Theory and Model

**Theory of Emergent Facilitation.** To begin, we describe how glassy dynamics may be understood as a hopping process in a potential energy landscape (PEL) (22–24). The PEL is a landscape in the high-dimensional configuration space composed of multiple local minima, where the energy-minimizing configurations are referred to as inherent states (ISs). The hopping process involves the system moving across different ISs in the configuration space, as illustrated in Fig. 1(a). Connecting two nearby ISs is a transition state, which the system can cross on an average hopping timescale $\tau_{hop}$ (5). Starting from an initial IS, previous work (5) indicates that the supercooled liquid experiences a bond-exchange event when crossing a transition state, as illustrated in Fig. 1(b). We refer to these bond-exchange events as excitations, and their formation reorganizes the surrounding medium via pure shear (5, 25). Within timescales of $O(\tau_{hop})$, the characteristics of excitations are insensitive to the choice of initial IS and so shifting the time origin allows us to observe similar excitations. The full relaxation process, however, operates on a timescale $\tau_{eq} \gg \tau_{hop}$. Therefore, a theory of glassy dynamics should be able to describe relaxation based on a series of excitations or bond-exchange events connecting nearby ISs.

Linear elasticity theory provides a mechanism for which multiple excitations mediate glassy relaxation dynamics in supercooled liquids. To see this, we examine the aftermath of an initial excitation in a supercooled liquid, i.e., when the system crosses the first transition state. In this case, the ensuing bond-exchange event leaves behind mechanical stresses with respect to the initial IS. The presence of these stresses impacts the formation of a new bond-exchange event through elastic interactions with the previous excitation, thus changing the cost to form new excitations. At very low temperatures, these elastic interactions strongly influence the overall energy barriers to relax the system via the creation of multiple excitations, and thus determine the most probable relaxation pathway of the system. Given a model for relaxation that captures these elastic interactions, we may then observe the emergence of facilitation and glassy dynamics.

A mathematical framework that can capture such a relaxation pathway is a Markov process (26). As applied to supercooled liquids, the Markov process models stochastic jumps between ISs, occurring on a timestep of $O(\tau_{hop})$ so that glassy behavior may emerge on timescales $\tau_{eq} \gg \tau_{hop}$; see Fig. 1(a) for illustration. At time $t = 0$, the system begins with an empty state with no excitations, representing an arbitrary IS, and excitations emerge as time progresses. Denoting a state via an index $\alpha$, we can write the master equation describing the time evolution of the probability $p_\alpha(t)$ to be in state $\alpha$ (26) as follows:

$$\frac{\mathrm{d}p_\alpha(t)}{\mathrm{d}t} = \sum_{\alpha'} w_{\alpha\alpha'} p_{\alpha'}(t) - w_{\alpha'\alpha} p_\alpha(t), \qquad [1]$$

where $w_{\alpha'\alpha}$ is the rate to transition from state $\alpha$ to $\alpha'$, with an initial condition as the empty state ($\alpha = 0$), i.e., $p_\alpha(0) = \delta_{\alpha 0}$.

The form of transition rates $w_{\alpha'\alpha}$ is given by transition state theory (TST) (27),

$$w_{\alpha'\alpha} = \nu_0 e^{-\beta \Delta F^\ddagger_{\alpha'\alpha}}, \qquad [2]$$

where $\nu_0$ is a frequency prefactor, which is assumed to be constant in this work, and $\Delta F^\ddagger_{\alpha'\alpha}$ is the free-energy barrier. From detailed balance, the potential energy difference $\Delta U_{\alpha'\alpha}$ between two ISs can also affect Eq. (2). However, we may argue that $\Delta U_{\alpha'\alpha} \approx 0$ when comparing energy differences $\Delta U^{IS}$ between ISs with that of the barriers $\Delta U^\ddagger$ connecting nearby ISs. On a timescale $\tau_{hop}$, for a system size $N$, we create $N_{exc}$-many excitations on average where $N_{exc} \sim Nc_\sigma$ with $c_\sigma$ the concentration of excitations (5). This implies that $\Delta U^\ddagger \sim N_{exc} J_\sigma \sim Ne^{-\beta J_\sigma} J_\sigma$. If we assume that $\Delta U^{IS}$ is proportional to the standard deviation in IS energies, then $\Delta U^{IS} \sim \sqrt{k_B T^2 c_v^{IS} N}$, with $c_v^{IS}$ being the IS heat capacity. In that case, $\Delta U^{IS}/\Delta U^\ddagger \sim \sqrt{k_B T^2 c_v^{IS} N}/(Ne^{-\beta J_\sigma} J_\sigma) \sim O\left(1/\sqrt{N}\right)$, which goes to zero in the thermodynamic limit. Thus, we may set $\Delta U_{\alpha'\alpha} \approx 0$, modeling a system that hops between energy wells of equal depths but with different energy barriers, which we now address in detail.

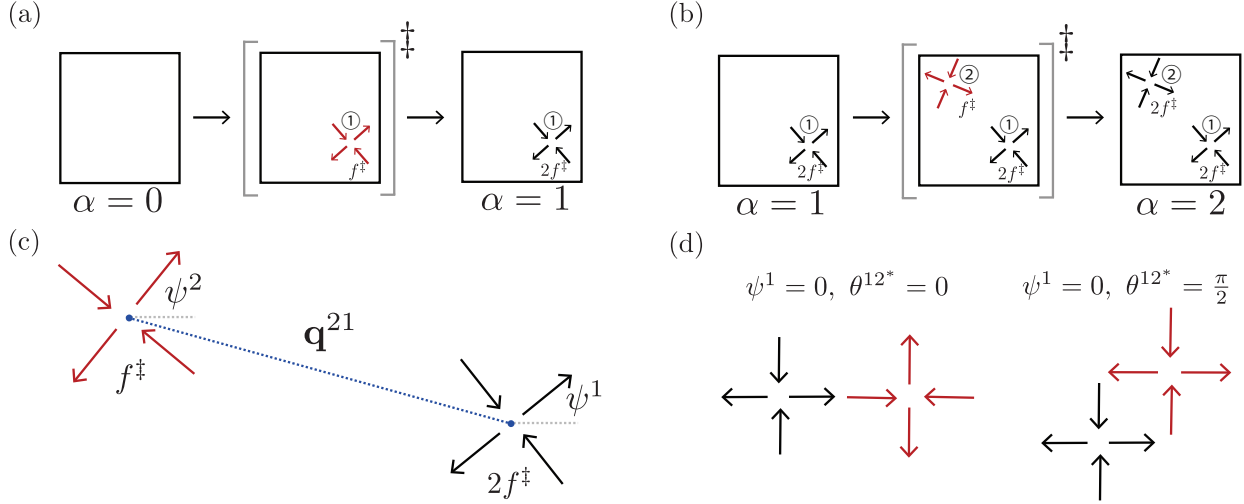

**Fig. 2.** (a) Schematic of a transition from an empty state to a state with one excitation. The transition state for creating excitation ① (red arrows) can be represented by a pair of force dipoles with magnitude $f^\ddagger$ leading to a bond-exchange event as in Fig. 1(b). Once the system completes its transition, excitation ① induces mechanical stresses in the medium with the same set of force dipoles (black arrows) but with magnitude $2f^\ddagger$, corresponding to the full bond-exchange event. (b) Transition from a state with one excitation to another with two excitations. The transition state for creating excitation ② (red arrows) involves elastic interactions with the mechanical stresses induced by the prior excitation ① (black arrows). Once the transition is complete, the system enters a new state with two excitations, each of which induces mechanical stresses corresponding to two pairs of force dipoles with magnitude $2f^\ddagger$ (black arrows). (c) A schematic of the force-dipole configuration describing the elastic interactions given by Eq. (12). (d) Various configurations minimizing the free-energy barriers associated with elastic interactions in Eqs. (11) and (12), indicating emergent facilitation at low temperatures.

Following previous works (5, 14), we use linear elasticity theory to compute the barrier $\Delta F^\ddagger_{\alpha'\alpha}$, which can be written as a superposition of two parts: (1) the cost to create an excitation, denoted as $J_\sigma$, and (2) the interaction of an excitation with the stresses $T^\alpha_{ij}$ present at the current state $\alpha$, with $i,j \in \{1,2\}$. The stress components $T^\alpha_{ij}$ build over time due to accumulated stresses induced by past excitations. In the framework of linear elasticity, the formula for $\Delta F^\ddagger_{\alpha'\alpha}$ can be written using the superposition principle as

$$\Delta F^\ddagger_{\alpha'\alpha} = J_\sigma + \int d^2\mathbf{x}\, T^\alpha_{ij} d^\ddagger_{ij} \quad [3]$$

where $d^\ddagger_{ij} = \epsilon^\ddagger_{ij} - \tfrac{1}{2}\epsilon^\ddagger_{kk}\delta_{ij}$ is the deviatoric part of the strain tensor $\epsilon^\ddagger_{ij}$ arising from a new bond-exchange event leading to the new state $\alpha'$. The energy cost $J_\sigma$ for a single excitation follows from our previous work (5) where we developed a theory of localized excitations in supercooled liquids, and is written as

$$J_\sigma = \frac{1}{2}\int d^2\mathbf{x}\, d^\ddagger_{ij} C^{\rm IS}_{ijkl} d^\ddagger_{kl}, \quad [4]$$

with $C^{\rm IS}_{ijkl}$ being the IS elasticity tensor, whose formula can be derived from first principles (5, 28, 29). Note that Equation (3) may be viewed as a modification of $J_\sigma$ by allowing it to change as excitations accumulate stresses over time.

With the general formula in Eq. (3), we now study in the following sections how different scenarios for bond-exchange events modify the overall energy barrier, and thereby give rise to facilitation.

**Empty State → One Excitation.** We begin with the simplest case, involving a transition from an initially empty state ($\alpha = 0$) to one with a single excitation ($\alpha' = 1$). In this case, the system experiences a bond-exchange event with zero initial stresses, i.e., $T^0_{ij} = 0$, and thus the free-energy barrier for this transition is $\Delta F^\ddagger_{10} = J_\sigma$. In 2D, we can compute $J_\sigma$ from Eq. (4) using the two nonzero components of the deviatoric strain tensor $d^\ddagger_{ij}$. In particular, given a symmetric tensor such as $d^\ddagger_{ij}$, we can decompose it in terms of its hydrostatic component $d^\ddagger_1 = -(d^\ddagger_{11} + d^\ddagger_{22})/\sqrt{2}$, which vanishes to zero by definition, and the deviatoric components $d^\ddagger_2 = (d^\ddagger_{22} - d^\ddagger_{11})/\sqrt{2}$ and $d^\ddagger_3 = \sqrt{2}d^\ddagger_{12}$. For an isotropic elastic system, Eq. (4) can then be rewritten as

$$J_\sigma = \int d^2\mathbf{x}\, G^{\rm IS}\left[\left(d^\ddagger_2\right)^2 + \left(d^\ddagger_3\right)^2\right], \quad [5]$$

where $G^{\rm IS}$ is the IS shear modulus (see Supplemental Material (SM) Sec. 1).

Within the linear elasticity theory, a pure-shear excitation can be represented by two pairs of opposing point forces, referred to as force dipoles (5, 30). The force dipoles are arranged so that an excitation induces localized pure shear corresponding to the bond-exchange event, as seen in Fig. 2(a). Each point force has a magnitude $f^\ddagger$ associated with the transition state, and every pair is separated by a distance twice the excitation radius $R_{\rm exc}$. Given the free-body diagram shown in Fig. 2(a), we can compute the deviatoric stress tensor $S^\ddagger_{ij} = T^\ddagger_{ij} - \tfrac{1}{2}\delta_{ij}T^\ddagger_{kk}$ associated with the force-dipole configuration in the transition state (5). By definition, the hydrostatic component $S^\ddagger_1 = 0$ and the remaining components can be written as

$$S^\ddagger_2(r,\theta;\psi) = \sqrt{2}f^\ddagger R_{\rm exc}\frac{(\nu^{\rm IS}+1)}{\pi r^2}\cos(4\theta - 2\psi), \quad [6]$$

$$S^\ddagger_3(r,\theta;\psi) = -\sqrt{2}f^\ddagger R_{\rm exc}\frac{(\nu^{\rm IS}+1)}{\pi r^2}\sin(4\theta - 2\psi), \quad [7]$$

where $\nu^{\rm IS}$ is the IS Poisson's ratio, $(r,\theta)$ are the polar coordinates and $\psi$ is the orientation angle of the excitation;

see SM Sec. 1 and Ref. (5) for derivations. The magnitude $f^{\ddagger}$ can be expressed as a function of the eigenstrain $\epsilon_c$, i.e., $f^{\ddagger} = \frac{\sqrt{2}\pi R_{\text{exc}} G^{\text{IS}} \epsilon_c}{1+\nu^{\text{IS}}}$ (5), where $\epsilon_c$ determines the extent of uniform inelastic deformation inside the excitation radius (31, 32). Using the formula for $f^{\ddagger}$, the constitutive relation $S_{ij}^{\ddagger} = 2G^{\text{IS}} d_{ij}^{\ddagger}$, Eqs. (6) and (7), we can write the corresponding deviatoric strain fields as

$$d_2^{\ddagger}(r,\theta;\psi) = \frac{\epsilon_c R_{\text{exc}}^2}{r^2} \cos(4\theta - 2\psi), \quad [8]$$

$$d_3^{\ddagger}(r,\theta;\psi) = -\frac{\epsilon_c R_{\text{exc}}^2}{r^2} \sin(4\theta - 2\psi), \quad [9]$$

Substituting Eqs. (8) and (9) into Eq. (5), we obtain

$$J_\sigma = G^{\text{IS}} \epsilon_c^2 \pi R_{\text{exc}}^2 \quad [10]$$

which indicates that $J_\sigma$ is the energy required to reorganize an elastic medium by shearing a small region of radius $R_{\text{exc}}$ with a uniform strain $\epsilon_c$.

In 2D systems, $\epsilon_c$ and $R_{\text{exc}}$ can be computed by associating the localized pure-shear deformation with a T1 transition event (5). During a T1 transition, particles participate in a bond-exchange event by pushing an initial bond connecting the nearest neighbors beyond a critical displacement $u^{\ddagger}$ corresponding to the transition state, leading to the formation of a new bond (Fig. 1(b)). The geometry of the particle configuration allows $\epsilon_c$ and $R_{\text{exc}}$ to be estimated as a function of $u^{\ddagger}$, the value of which is further determined from the local structure, i.e., the inherent-state RDF (5). Together with $G^{\text{IS}}$, $J_\sigma$ can thus be computed using only static equilibrium properties. *

**One Excitation → Two Excitations.** Once the first excitation is created, the bond-exchange event completes its full transition, imposing mechanical stresses $T_{ij}^1$ onto the surrounding medium. These stresses arising from the full bond-exchange event involve double the critical displacement magnitude $u^{\ddagger}$ at the transition state and therefore correspond to imposing a pair of force dipoles of magnitude $2f^{\ddagger}$ (Fig. 2(b)). The next bond-exchange event, which corresponds to a new pair of force dipoles with magnitude $f^{\ddagger}$, will be influenced by the imposed stresses $T_{ij}^1$. Consequently, the next transition from a state with one excitation ($\alpha = 1$) to a state with two excitations ($\alpha = 2$) involves a nonzero interaction term in the energy barrier. As shown in the free-body diagram in Fig. 2(b), the transition state now consists of a pair of force dipoles with magnitude $f^{\ddagger}$, which is the new excitation, and another pair of force dipoles from the previous excitation with magnitude $2f^{\ddagger}$ that exerts the stress field $T_{ij}^1$. In the context of linear elasticity theory, this results in a modified energy barrier containing long-range elastic interactions arising from Eq. (3):

$$\Delta F_{21}^{\ddagger} = J_\sigma + v_{\text{int}}^{21}. \quad [11]$$

where $v_{\text{int}}^{21}$ is the elastic interaction,

$$v_{\text{int}}^{21} = \frac{\kappa J_\sigma}{(\tilde{q}^{21})^2} \cos\left(2\psi^2 + 2\psi^1 - 4\theta^{21}\right) \quad [12]$$

where $\psi^2$ and $\psi^1$ are the orientation angles of excitations ② and ①, respectively, $\tilde{q}^{21} = q^{21}/(2R_{\text{exc}})$ is the pair distance $q^{21}$ between the two excitations normalized by the diameter of the

---
*For more details regarding the derivation of $J_\sigma$ and testing the formula on MD simulations of many polydisperse glass formers, see Ref. (5).

excitation $2R_{\text{exc}}$, $\theta^{21}$ is the polar angle of the pair distance vector $\mathbf{q}^{21}$, and $\kappa = \frac{2}{1+\nu^{\text{IS}}}$ adjusts the interaction strength relative to $J_\sigma$; see Fig. 2(c) for a diagram and SM Sec. 2 for the derivation.

The elastic interaction in Eq. (11) is the key to emergent facilitation as temperature $T \to 0$. To see this, we look for the most probable excitation at low temperatures, which corresponds to minimizing the modified energy barrier in Eq. (11). Let $\tilde{\mathbf{q}}^{21*} = \tilde{q}^{21*}[\cos\theta^{21*}, \sin\theta^{21*}]$ be the energy-minimizing position of excitation ② relative to excitation ① and $\psi^{2*}$ be the corresponding orientation angle of excitation ②. The equations defining the energy minima can be written as follows (SM Sec. 3):

$$\psi^{2*} = 2\theta^{21*} + \pi\left(n + \frac{1}{2}\right) - \psi^1, \quad \tilde{q}^{21*} = 1 \quad [13]$$

where $n$ is an integer. For every choice of polar angle $\theta^{21*}$, Eq. (13) guarantees an orientation angle $\psi^{2*}$ that allows the two excitations to be as close as possible; see Fig. 2(d) for examples of energy-minimizing configurations. On the other hand, the probability of finding a new excitation at high temperatures is uniform in space since such probability is proportional to $e^{-\beta v_{\text{int}}^{21}} \to 1$ as $T \to \infty$. Thus, we see how the strength of elastic interactions grows from high temperatures, where the next excitation can be formed anywhere in space, to low temperatures, where dynamical facilitation emerges in the sense that an excitation leads to further excitations nearby.

**Revertibility of Excitations.** In supercooled liquids, we often encounter bond-exchange events that either (1) reverse previous bond-exchange events, thus returning the system to its previous state or (2) emerge at the same spot as where the previous bond-exchange events are, thus taking the system to a new state. The latter case amounts to reorganizing the region that has already relaxed (10). In the Markov process, however, both cases correspond to reverting existing excitations. Computing the free-energy barrier for such excitations can also be done via free-body diagrams. For instance, suppose we study the first case corresponding to the backward transition in Fig. 2(a), i.e., hopping from state $\alpha = 2$ to $\alpha' = 1$ (Fig. 3(a)), which models the event where a new bond-exchange event reverts and brings the system back to its previous state. This reversal event can be modeled via a free-body diagram shown in Fig. 3(b), where the reverse excitation is put on top of the second excitation to produce a pair of force dipoles with magnitude $f^{\ddagger}$. Since the transition state we obtain is identical to the one from the forward transition (see Fig. 2(a)), the barrier for this transition is $\Delta F_{21}^{\ddagger} = \Delta F_{12}^{\ddagger}$.

Next, let us consider excitations that occur at the same spot where the previous excitations are and yet bring the system to a new state. This corresponds to studying a series of two transitions depicted in Fig. 3(c). Starting with an initial state $\alpha = 2$ with excitations ① and ②, suppose the system creates a new excitation ③ taking it to state $\alpha = 3$. The free-energy barrier for this transition from Eq. (3) is $\Delta F_{32}^{\ddagger} = J_\sigma + v_{\text{int}}^{31} + v_{\text{int}}^{32}$ according to the free-body diagram. Suppose the next hopping event creates further reorganization at excitation ①, which amounts to reverting excitation ①. Such a transition leads to a final state $\alpha = 4$ that is different from the initial state $\alpha = 2$, despite having the same number of excitations. The transition state for this event amounts to studying the free-body diagram obtained by adding a reverse excitation on excitation ① shown in Fig. 3(c), from which we can deduce that the energy barrier is $\Delta F_{43}^\ddagger = J_\sigma + v_{\text{int}}^{12} + v_{\text{int}}^{13}$. Note, however, that the backward version of this transition, where the system creates excitation ① in the presence of excitations ② and ③, also leads to the same transition state, and thus $\Delta F_{43}^\ddagger = \Delta F_{34}^\ddagger$.

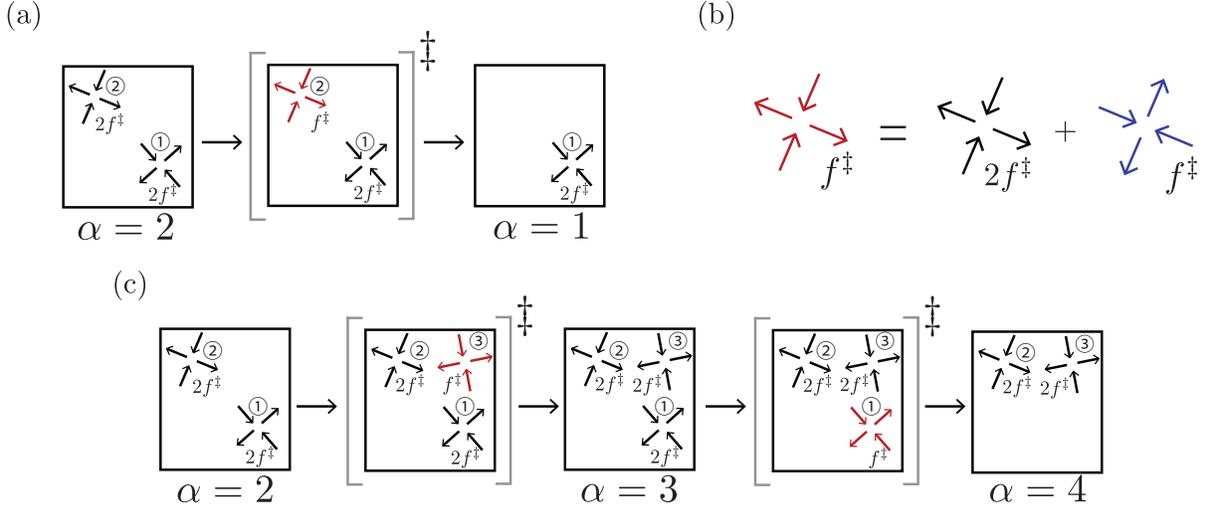

**Fig. 3.** Schematic of transitions indicating revertibility of excitations. (a) A transition depicting the backward version of Fig. 2(a), where the bond exchange event reverts, taking the system back to its previous state. The transition state for this event amounts to reducing the force magnitude of excitation ② from $2f^\ddagger$ to $f^\ddagger$. (b) The free-body diagram describing the transition state for the event in (a) amounts to adding a pair of force dipoles in the reverse direction (blue arrows). (c) A series of excitations, beginning from the final state of Fig. 2(b), that takes the system from a state with two excitations to another distinct inherent state with the same number of excitations. The first transition involves creating excitation ③ while the second transition reverts excitation ①, indicating the reorganization of regions that have already relaxed.

**General Forward and Backward Transitions.** The use of free-body diagrams involving pairs of force dipoles leads us to conclude that given a state $\alpha$ with $n$-many excitations, the free-energy barrier to add a new excitation (labeled $\mu$) leading to a new state $\alpha'$ is via superposition principle

$$\Delta F_{\alpha'\alpha}^\ddagger = J_\sigma + \sum_{\mu' \neq \mu} v_{\text{int}}^{\mu\mu'}, \quad [14]$$

$$v_{\text{int}}^{\mu\mu'} = \frac{\kappa J_\sigma}{(\tilde{q}^{\mu\mu'})^2} \cos\left(2\psi^\mu + 2\psi^{\mu'} - 4\theta^{\mu\mu'}\right). \quad [15]$$

Furthermore, the backward transition involving reverting the excitation $\mu$ passes through the same transition state as the forward one, leading to

$$\Delta F_{\alpha\alpha'} = \Delta F_{\alpha'\alpha}, \quad [16]$$

which is also consistent with the equal energy well picture.

**Model Implementation.** The theory described until now corresponds to excitations that appear anywhere in continuous real space. However, known algorithms for simulating Markov processes are well suited in a discrete real space, e.g., the kinetic Monte Carlo (kMC) algorithm (33). To that end, we triangulate the 2D continuous real space with a lattice lengthscale $\ell_{\text{d}}$, where excitations occupy $N_\ell$-many lattice sites, each of which is labeled $\mu$ so that the set of lattice site positions is $\{\mathbf{x}_\mu\}_{\mu=1}^{N_\ell}$. This new lengthscale is typically on the order of the excitation diameter, i.e., $\ell_{\text{d}} \sim O(2R_{\text{exc}})$, and thus an occupied lattice site approximately represents an excitation with excluded volume. It can also be used to satisfy the non-overlap condition of the elastic interaction between two excitations, which can now be rewritten as

$$v_{\text{int}}^{\mu\mu'} = \frac{\hat{\kappa} J_\sigma}{(\hat{q}^{\mu\mu'})^2} \cos\left(2\psi^\mu + 2\psi^{\mu'} - 4\theta^{\mu\mu'}\right) \quad [17]$$

where $\hat{\kappa} = \kappa(2R_{\text{exc}}/\ell_{\text{d}})^2$ is the new effective interaction strength and $\hat{q}^{\mu\mu'} = q^{\mu\mu'}/\ell_{\text{d}}$. Due to the spatial discretization, the transition rates are also modified into

$$\hat{w}_{\alpha'\alpha} = A\nu_0 e^{-\beta\Delta F_{\alpha'\alpha}^\ddagger}, \quad [18]$$

where $A$ is an additional prefactor that accounts for the translational entropy of creating an excitation. The prefactor $A = \frac{\sqrt{3}}{2}(\ell_{\text{d}}/\sigma_{\text{exc}})^2$ when we insert a new excitation and $A = 1$ otherwise; see SM Sec. 4.1 for derivations.

Given Eq. (18), we use the kMC algorithm (33) to simulate the Markov process. As applied to our model, the kMC algorithm generates stochastic jumps of the Markov process, creating a trajectory of excitations that uses an empty state as the initial condition. At every $k$-th jump, we perform two steps: (1) sample a new state $\alpha_{k+1}$ based on the discrete probability distribution $p(\alpha_{k+1}; \alpha_k) = \hat{w}_{\alpha_{k+1}\alpha_k}/\hat{W}_{\alpha_k}$, where $\hat{W}_{\alpha_k} = \sum_{\alpha'} \hat{w}_{\alpha'\alpha_k}$ is the total transition rate, and (2) sample a new waiting time $\tau_k$ based on an exponential distribution, with the rate parameter being $\hat{W}_{\alpha_k}$. The output of the kMC algorithm is a relaxation pathway containing a series of states and arrival times $\{\alpha_k, t_k\}$, which we can use to probe relaxation behaviors arising from the model. For more details on implementing the kMC algorithm to our model of elastically interacting excitations, see SM Sec. 4.

## Results

We now discuss several results obtained from the kMC simulations, demonstrating the ability of the model to reproduce key elements of glassy dynamics. Unless otherwise noted, all simulation results use the lattice lengthscale as $\ell_{\text{d}}/(2R_{\text{exc}}) = 4/3$, and the system size as $L = 15\ell_\mathrm{d}$. Details on the model parameters used in the kMC simulations can be found in SM Sec. 4.3.

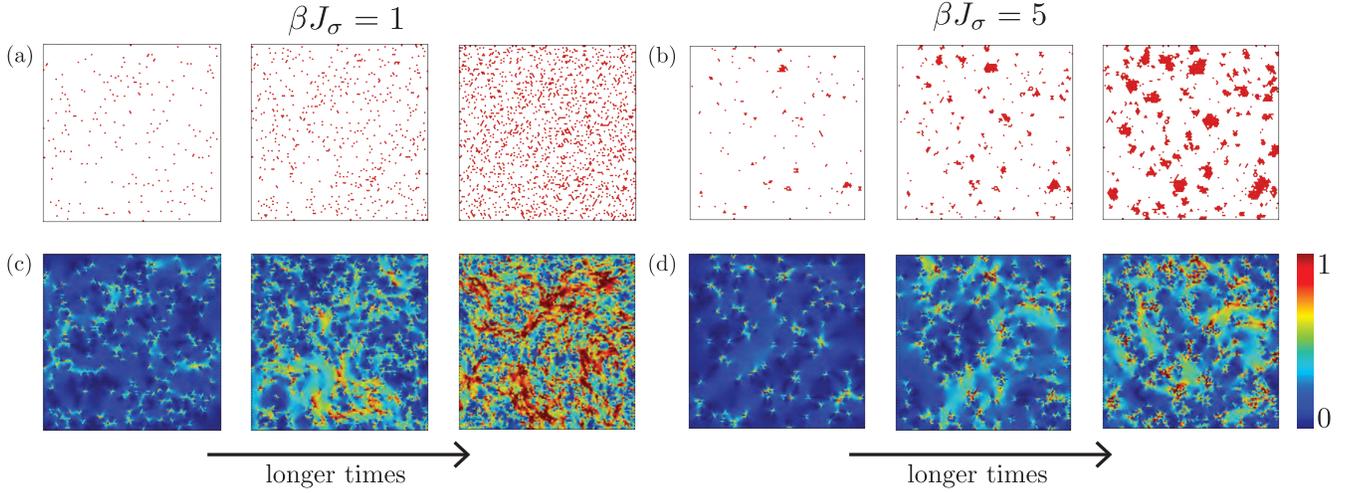

**Fig. 4.** Snapshots of the persistence variable tracking the activity of a region with time at high (a) and low (b) temperatures corresponding to $\beta J_\sigma = 1$ and $\beta J_\sigma = 5$, respectively. At high temperatures, we see a Poisson point process where excitations arrive randomly in space and time. However, at low temperatures, we see the emergence of dynamical facilitation with clustered mobile regions that grow in time. Also shown are the displacement fields for the corresponding high (c) and low (d) temperatures. Simulations correspond to a system size of $L = 100\ell_\mathrm{d}$ and an interaction strength $\kappa = 1/2$. The color code in (c) and (d) indicate the magnitude of particle displacements relative to particle diameter $\sigma$.

**Emergence of Dynamical Facilitation.** We begin by understanding how dynamics within the model proceed in space and time. To this end, we use the persistence variable $p_\mu(t)$ and the displacement field $\mathbf{u}(\mathbf{x}, t)$. The persistence variable $p_\mu(t)$ tracks the activity of a lattice site; $p_\mu(t) = 1$ if an excitation has arrived at least once at lattice position $\mathbf{x}_\mu$ before time $t$, and zero otherwise. The displacement field $\mathbf{u}(\mathbf{x}, t)$ can be computed given the set of excitation positions $\{\mathbf{q}^\mu\}$, orientation angles $\{\psi^\mu\}$, and arrival times $\{t^\mu\}$ as follows:

$$\mathbf{u}(\mathbf{x}, t) = \sum_{\mu=1}^{N_\mathrm{exc}(t)} \mathbf{u}^\mathrm{exc}(\mathbf{x} - \mathbf{q}^\mu; \psi^\mu) \Theta(t^\mu - t) \quad [19]$$

where $N_\mathrm{exc}(t)$ is the total number of excitations at time $t$, $\mathbf{u}^\mathrm{exc}(\mathbf{x} - \mathbf{q}^\mu; \psi^\mu)$ is the elastic displacement field associated with a single excitation and $\Theta(s)$ is the Heaviside function; see SM Sec. 5 for a derivation of $\mathbf{u}^\mathrm{exc}$.

Figures 4(a)&(b) show sample trajectories visualized by the persistence variable $p_\mu(t)$ and displacement field $\mathbf{u}(\mathbf{x}, t)$ at two different temperatures. At high temperatures ($\beta J_\sigma = 1$, Fig. 4(a)), excitations or bond-exchanges arise as random events in space—a behavior closely resembling a Poisson point process (26)—and continue to do so as time progresses (see SM Movie S.1). In contrast, as the temperature is lowered as shown in Fig. 4(b) for $\beta J_\sigma = 5$, dynamics proceed in two stages. In the first stage, excitations sparsely populate the system, similar to the high-temperature process. This is followed by the next stage, where excitations occur near the previous excitations, leading to spatial clustering of dynamics. These clusters then grow in size as more excitations arrive over time (see SM Movie S.2). The same is reflected in the displacement behaviors, where at high temperatures the displacement fields develop in a spatially uncorrelated manner (Fig. 4(c); see SM Movie S.3). In contrast, we see the displacement fields at low temperatures

are increasingly concentrated in regions of the clusters of excitations (Fig. 4(d)); see also SM Movie S.4. Furthermore, we also observe the development of branched-liked patterns in the displacement field at such low temperatures, a typical signature of dynamical heterogeneity as seen in MD simulations (2, 3).

The two-stage behavior at low temperatures can be explained as follows. At short times, excitations arrive at a rate proportional to $e^{-\beta J_\sigma}$ with barrier $J_\sigma$, and thus are in small concentrations at low temperatures. Since excitations interact very little at such low concentrations, they essentially behave as a Poisson point process. As time progresses, however, the number of excitations increases, and thus elastic interactions grow stronger. In particular, these elastic interactions attract new excitations to the previous ones, resulting in the clustering of excitations and subsequent spreading of high-mobile regions—all of which are signatures of emergent facilitation. Note that these emergent behaviors are consistent with recent MD simulation studies (10), where tracking bond-exchange events in a supercooled polydisperse fluid reveals the clustering and spreading of high-mobility regions.

**Exponential to Stretched Exponential Relaxation Behaviors.** The emergence of dynamical facilitation also quantitatively impacts relaxation dynamics. To explore this, we compute the bond-order autocorrelation function $C_\mathrm{b}(t)$ and self-intermediate scattering function $F_\mathrm{s}(\mathbf{k}, t)$, both of which are common measures for extracting relaxation timescales. The bond-order function $C_\mathrm{b}(t)$ encodes rotational relaxation of the system, and can be computed as

$$C_\mathrm{b}(t) := \frac{1}{N_\ell} \sum_{\mu=1}^{N_\ell} \mathbb{E}\left[e^{i6\theta(\mathbf{x}_\mu, t)}\right] \quad [20]$$

$$\theta(\mathbf{x}, t) = \sum_{\mu=1}^{N_\mathrm{exc}(t)} \theta_\mathrm{exc}(\mathbf{x} - \mathbf{q}^\mu; \psi^\mu) \Theta(t^\mu - t) \quad [21]$$

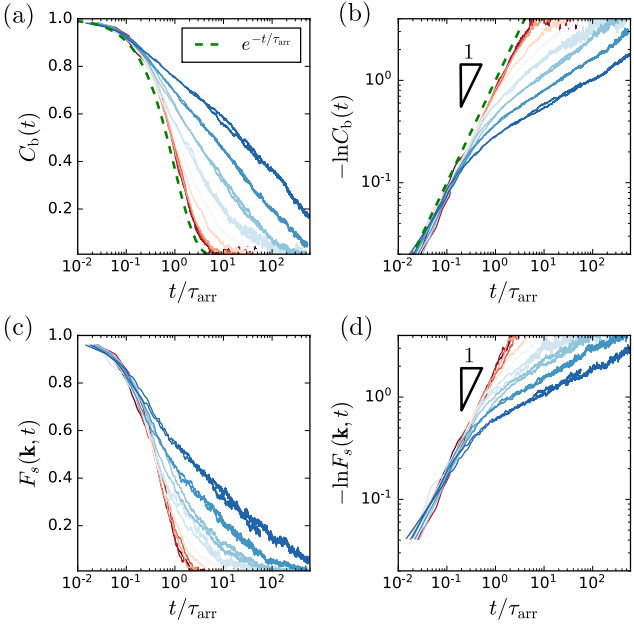

**Fig. 5.** Relaxation functions computed from kMC simulations for $L = 15\ell_d$ and $\kappa = 1/2$: (a,b) bond-order autocorrelation function $C_b(t)$ and (c,d) the self-intermediate scattering function $F_s(\mathbf{k}, t)$. From red to blue, the inverse temperature $\beta J_\sigma \in [0, 3.86]$. Both relaxation functions exhibit a crossover from exponential to stretched exponential relaxation behaviors from high to low temperatures. For every temperature, kMC simulations are run for different decades of time, producing multiple curves and allowing us to resolve the relaxation process at a computationally tractable sampling rate. We show all curves for completeness even though the longest run is sufficient.

where $\mathbb{E}[\ldots]$ is an ensemble average over trajectories, and $\theta(\mathbf{x}, t)$ measures the rotational strain induced by excitations with $\theta_{\text{exc}} = \frac{1}{2}\left(\partial_y u_x^{\text{exc}} - \partial_x u_y^{\text{exc}}\right)$ and $u_x^{\text{exc}}$ and $u_y^{\text{exc}}$ being the $x$- and $y$-component, respectively, of $\mathbf{u}^{\text{exc}}$. Figures 5(a)&(b) show $C_b(t)$ for varying temperatures, with respect to an Arrhenius time scale $\tau_{\text{arr}} \sim e^{\beta J_\sigma}$, corresponding to the average hopping time between inherent states ($\tau_{\text{hop}}$) or the average time for the arrival of an excitation. The bond-order function decays exponentially at high temperatures at all times. In contrast, we see the emergence of a long-time stretched exponential decay at low temperatures, i.e., $C_b(t) \sim e^{-(t/\tau_b)^a}$, where $\tau_b$ is the temperature-dependent bond-relaxation time and the stretching exponent $a \approx 0.33 - 0.38$.

Similar behaviors are also observed in the self-intermediate scattering function $F_s(\mathbf{k}, t)$, which is computed as

$$F_s(\mathbf{k}, t) := \frac{1}{N_\ell} \sum_{\mu=1}^{N_\ell} \mathbb{E}\left[e^{i\mathbf{k} \cdot \mathbf{u}(\mathbf{x}_\mu, t)}\right], \quad [22]$$

where we choose $|\mathbf{k}| = 2\pi/\sigma$ with $\sigma$ being a characteristic particle diameter. Here, Figs. 5(c)&(d) again show a crossover from a high-temperature exponential decay to a low-temperature stretched exponential decay, i.e., $F_s(\mathbf{k}, t) \sim e^{-(t/\tau_s)^a}$ for $|\mathbf{k}| = 2\pi/\sigma$, with $\tau_s$ the structural relaxation time and the corresponding stretching exponent in the same range as that of $C_b(t)$. The existence of a stretched exponential decay in both $C_b(t)$ and $F_s(\mathbf{k}, t)$ at low temperatures is a vital signature of glassy relaxation and has been observed in numerous experiments and MD simulations (1).

Understanding the crossover from an exponential to a stretched exponential decay in $C_b(t)$ goes back to the emergence of dynamical facilitation. At short timescales, excitations arrive as a Poisson point process, with negligible elastic interactions. The characteristic time $\tau_{\text{arr}}$ for this process is the timescale needed for the concentration of excitations $c_\sigma$ to reach $c_\sigma \sim e^{-\beta J_\sigma}$. At high temperatures, $c_\sigma \to 1$ on timescales of $\tau_{\text{arr}}$, which implies that the initial Poisson point process can lead to the accumulation of excitations or bond-exchange events that relax the entire system without facilitation. In this case, for independent excitations, we may use the formalism of random point processes (26), which results in the bond-relaxation function as (34):

$$C_b(t) = e^{-t/\tau_{\text{arr}}} \quad [23]$$

$$\tau_{\text{arr}} = \frac{\sqrt{2}(1 + \nu^{\text{IS}})}{6\pi^2 \epsilon_c \nu_0} e^{\beta J_\sigma} . \quad [24]$$

Figures 5(a)-(b) show that Eq. (23) is in good quantitative agreement with the kMC simulations at sufficiently high temperatures. As the temperature decreases, the concentration at the characteristic Arrhenius timescale $c_\sigma \to 0$ and, consequently, a Poisson process is insufficient to fully relax the system. Here, the relaxation function is consistent with Eq. (23) only for short timescales when $t < \tau_{\text{arr}}$. For timescales beyond $\tau_{\text{arr}}$, facilitation emerges from the elastic interactions and is thus responsible for any deviation from the Poissonian behavior. We do not yet have an analytical theory to derive the form of the stretched exponential decay from the model, and leave such an endeavor to future work.

**Arrhenius to Super-Arrhenius Relaxation Timescales.** The impact of emergent facilitation can also be seen in how relaxation timescales increase with decreasing temperature. To this end, we define the bond-orientational and structural relaxation times $\tau_b$ and $\tau_s$, respectively, as $C_b(\tau_b) = F_s(\mathbf{k}, \tau_s) = 0.1$. Figures 6(a)&(c) show the Arrhenius plot of relaxation timescales $\tau_b$ and $\tau_s$ for various interaction strengths $\kappa$. We observe a crossover from the high-temperature Arrhenius behavior towards the low-temperature super-Arrhenius behavior, with larger $\kappa$ leading to a more dramatic increase in relaxation timescales. The Arrhenius behavior arises from the independent and Poissonian nature of the arrival of the excitation events leading to $\tau_b, \tau_s \sim \tau_{\text{arr}}$.

The super-Arrhenius behavior arises primarily from the revertibility of excitations during the stage of emergent facilitation, wherein the system constantly destroys and forms new excitations in the same spot where the initial excitation is (see SM Movie S.5). This amounts to repeated reorganization of the regions that have already relaxed, traversing the system through new inherent states in the configuration space. These repeated bond-exchange events lead to slow growth of the cluster of excitations and therefore slow coarsening of the highly mobile regions, resulting in the dramatic slowdown of the relaxation process compared to the high-temperature Arrhenius regime. This observation can be further tested in the model, wherein we switch off the ability of the system to revert the excitations. In this case, facilitation still persists at low temperatures, where excitations arrive near previous excitations due to elastic interactions. However, the coarsening of the highly mobile regions occurs via a growth process proportional to the barrier $J_\sigma$, resulting in a relaxation time proportional to the Arrhenius time scale $\tau_{arr} \sim e^{\beta J_\sigma}$ even at low temperatures (see SM Fig. S.10). Note also the change in the nature of the relaxation function $C_b(t)$ from the stretched-exponential to an exponential decay (SM Fig. S.10). This shows that the super-Arrhenius and stretched-exponential behaviors in supercooled liquids result from an exploration of the high-dimensional configuration space, due to repeated reorganization of previously relaxed regions during the facilitation stage.

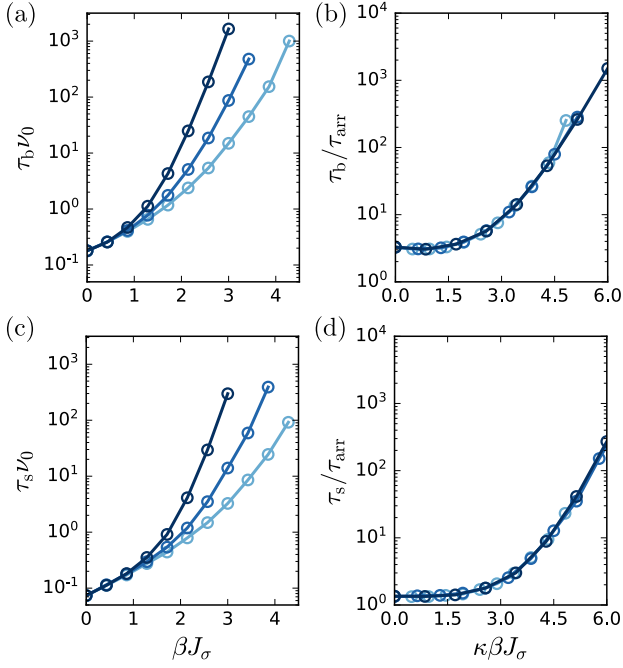

**Fig. 6.** The relaxation timescales $\tau_b$ (a) and $\tau_s$ (c) obtained from obtained from $C_b(t)$ and $F_s(\mathbf{k}, t)$, respectively, for different interaction strengths $\kappa \in [3/8, 1/2, 2/3]$ with $L = 15\ell_d$. The color code from light blue to dark blue corresponds to an increase in $\kappa$. Collapse of the relaxation timescales $\tau_b$ (c) and $\tau_s$ (d) upon an appropriate rescaling indicates super-Arrhenius behaviors are mediated by elastic interactions.

Furthermore, Figs. 6(b)&(d) show the universal collapse of the relaxation behaviors, by rescaling the relaxation times with respect to $\tau_{arr}$ and the dimensionless inverse temperature with $\kappa$. This shows that the deviation from the Arrhenius behaviors is mediated by the elastic interactions, indicating that different materials exhibit varying degrees of super-Arrhenius relaxation corresponding to their elastic interaction strengths. Such a collapse is another signature of glassy dynamics, which can be found in most supercooled liquids (1, 35).

**Finite Size Effects.** In 2D, elastic fields cause finite-size effects that substantially affect relaxation. It is now well-known that Mermin-Wagner fluctuations, which arise from the elastic vibrational motion of supercooled liquids, lead to a system-size dependence in the relaxation dynamics (14, 36–39). Here, we demonstrate how excitations can also result in additional finite-size effects and temporarily postpone the discussion of the impact of vibrational motion. Figure 7 shows relaxation times $\tau_b$ and $\tau_s$ for various system sizes, by again using the definition $C_b(\tau_b) = F_s(\mathbf{k}, \tau_s) = 0.1$. Here, we find that $\tau_b$ is independent of system size while $\tau_s$ decreases as system size increases, consistent with relaxation time observations in MD simulations of 2D model glass formers (36).

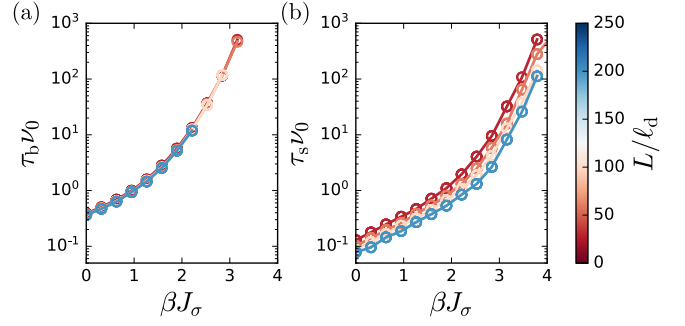

**Fig. 7.** Relaxation times $\tau_b$ (a) and $\tau_s$ (b) obtained from $C_b(t)$ and $F_s(\mathbf{k}, t)$, respectively, for various system sizes. Here, $\kappa = 1/2$ and $L$ is the linear system size, measured in units of lattice spacing length. The color code from red to blue corresponds to an increase in the system size, with $L \in \{30\ell_d, 60\ell_d, 100\ell_d, 200\ell_d\}$. There exist no finite-size effects in the bond-relaxation times, while the structural relaxation times show significant system-size effects and decrease as the system size increases.

At present, understanding the finite-size effects analytically is only possible in the high-temperature limit (34). Beginning with $\tau_b$, we know from Eq. (24) that $\lim_{T \to \infty} \tau_b = \tau_{arr}$, which is independent of system size. This result can be understood when computing $\tau_b$ from a Poisson point process at high temperatures, which to leading order for small rotational strains yields $\tau_b^{-1} \sim \int d^2\mathbf{x}\, \theta_{exc}^2$ (34). Since the excitation rotational strain $\theta_{exc}$ decays as $1/r^2$ (SM Sec. 5), we have $\tau_b^{-1} \sim \int_{\ell_d}^{L} \frac{dr\, r}{r^4} \sim O(1)$, indicating that $\theta_{exc}$ decays sufficiently fast that no finite-size effects can emerge. On the other hand, to leading order for small displacements at high temperatures, $\tau_s^{-1} \sim \int d^2\mathbf{x}\, (\mathbf{k} \cdot \mathbf{u}_{exc})^2$ (34). The excitation displacement field $\mathbf{u}_{exc}$ is long-ranged and decays as $1/r$ (SM Sec. 5), which results in $\tau_s^{-1} \sim \int_{\ell_d}^{L} \frac{dr\, r}{r^2} \sim O(\ln L)$, thus yielding a finite-size dependence analogous to Mermin-Wagner fluctuations where the relaxation proceeds faster by increasing system size. Indeed, applying the logarithmic finite-size scaling to Fig. 7(b) leads to a universal collapse of $\tau_s$ for all system sizes in the Arrhenius regime when $\beta J_\sigma \leq 2$ (see SM Fig. S.11). These finite-size effects produced by the model further support the standard practice in studying 2D glass formers, where $C_b(t)$ is the preferable relaxation function to compute compared to $F_s(\mathbf{k}, t)$ (10, 36, 39).

**Crossover Behaviors in the Mean Squared Displacement.** In addition to the relaxation functions, glassy dynamics also display their characteristic behavior in the mean-squared displacement (MSD). To that end, we compute an MSD by spatially averaging the square of the elastic displacement field, i.e.,

$$\text{MSD}^{el}(t) = \frac{1}{N_\ell} \sum_{\mu=1}^{N_\ell} \mathbb{E}\left[|\mathbf{u}(\mathbf{x}_\mu, t)|^2\right]. \quad [25]$$

We shall refer to Eq. (25) as the elastic MSD, which is shown for three different temperatures in Figure 8(a). At short times, the MSD is linear in time and collapses when rescaled with the Arrhenius timescale $\tau_{arr}$. As mentioned before, early-time dynamics proceed as a Poisson point process, which leads to

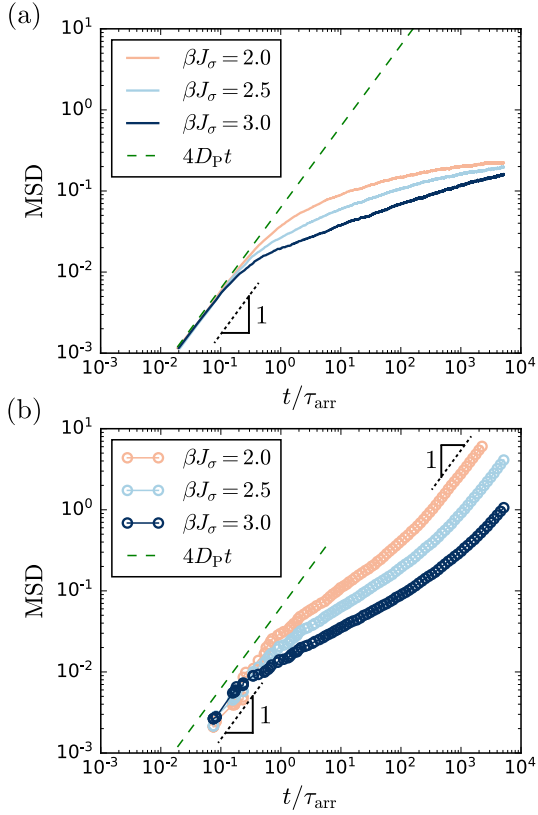

**Fig. 8.** The elastic MSD (a) and probe MSD (b) from inherent-state dynamics obtained at three different temperatures, where $\kappa = 2/3$, and $L = 15\ell_d$. The elastic MSD displays short-time diffusion-like behavior arising from excitations arriving as random events in the Arrhenius time scale ($\tau_{\rm arr} \sim \exp(-\beta J_\sigma)$). The behavior crosses over into subdiffusion at intermediate time scales ($t > \tau_{\rm arr}$), also coinciding with the emergence of dynamical facilitation. The probe MSD in (b), which requires resetting the relaxation process, shows the emergence of long-time diffusive behavior around the relaxation times.

the following short-time limit of Eq. (25):

$$\lim_{t \to 0} {\rm MSD}^{\rm el}(t) = 4 D_{\rm P} t, \qquad [26]$$

$$D_{\rm P} = \frac{\nu_0 (\pi u^\ddagger)^2}{8} \left(5 + \left(\nu^{\rm IS}\right)^2 - 2\nu^{\rm IS}\right) \left(\ln \frac{L}{R_{\rm exc}} + \frac{1}{2}\right) e^{-\beta J_\sigma}, \qquad [27]$$

which is in quantitative agreement with the kMC simulations (Fig. 8(a)). Note that the early-time dynamics are not ballistic as is commonly observed in realistic MSD, since the model only predicts the behaviors associated with inherent state transitions and therefore describes inherent-state dynamics.

Once the system passes the Poissonian regime at $t \sim \tau_{\rm arr}$, a subdiffusive regime appears coinciding with the emergence of the facilitation process, where ${\rm MSD} \sim t^{\gamma(T)}$ with the exponent $\gamma(T) < 1$. While we do not have an analytical theory for the exponent, the subdiffusion behavior can be qualitatively understood as follows. Physically, we know that subdiffusion may arise from anti-correlations in particle displacements, i.e., a forward motion is followed with high probability by a backward motion and vice versa (40). In the facilitation regime, a mechanism for such anticorrelation exists from the excitations that are constantly reforming within the same regions in space. It is plausible this process may result in elastic restorative forces that oppose the previous motion, likely causing the anticorrelations leading to subdiffusive MSD—a possibility that is left to be explored in future work. At longer times, however, the elastic MSD saturates to a plateau, unlike the diffusion behavior expected in an MSD. Such behavior emerges as a by-product of the model construction; at long times, excitations occupy almost all lattice sites and cannot accumulate further to relax the system beyond the relaxation timescale $\tau_{\rm b}$ when $C_{\rm b}(\tau_{\rm b}) \approx 0.1$. Thus, one cannot observe the long-time diffusive behavior, i.e., a post-relaxation process that emerges beyond $\tau_{\rm b}$, from computing just the elastic MSD.

Despite the model limitations, we may still study the long-time diffusive limit within the same model by constructing a special probe. In particular, we imagine a particle probe that can diffuse through the system via the following steps:

1. A probe starts at a reference position $\mathbf{R}_0$ while the system starts at an empty state at time $t = t_0$.

2. For every $k$-th step of the kMC simulation, the probe gets convected by the excitation displacement field, i.e.,

$$\mathbf{R}_k = \mathbf{R}_0 + \mathbf{u}(\mathbf{x} = \mathbf{R}_0, t_k). \qquad [28]$$

3. However, if an excitation arrives at a lattice site closest to $\mathbf{R}_0$ at the $k$-th step, reset the reference position $\mathbf{R}_0 \to \mathbf{R}_k$ and the system state to an empty state wiping out all the current excitations, restarting the relaxation process while keeping the time $t = t_k$.

4. Repeat step 2.

Step 3 is crucial as it allows the probe to reset its initial reference position and continue building its trajectory beyond the relaxation timescale; see SM Movie S.6 for an example trajectory of a probe particle at low temperatures ($\beta J_\sigma = 3.0$). We note that the average resetting time in Step 3 corresponds to a time scale for any arbitrary region to undergo a bond-exchange event. This resetting time also provides another measure of the relaxation time that we find commensurate with the earlier definition for $\tau_{\rm b}$.

The probe MSD is then computed from an ensemble average over trajectories as

$$\rm{MSD}^{\rm pr}(t) = \mathbb{E}\left[|\mathbf{R}(t)|^2\right] \qquad [29]$$

where $\mathbf{R}(t) = \mathbf{R}_k$ when the time $t$ falls within the arrival time range $t_k \leq t < t_{k+1}$. Figure 8(b) shows the probe MSD for three different temperatures consistent with the elastic MSD at short and intermediate time scales. Note the clear nature of the emergent subdiffusive behaviors in the probe MSD, scaling as $t^{\gamma(T)}$. We also see the emergence of a second cross-over from sub-diffusive to diffusive behaviors around the relaxation times, indicating that the long-time diffusive behavior is a post-relaxation process. In conclusion, the probe MSD transitions from Poissonian diffusion → subdiffusion → the long-time diffusion. These behaviors predicted by the model are consistent with recent works (10, 41), which, through MD simulations, have demonstrated the existence of both the Poissonian and subdiffusive behavior in the inherent-state dynamics, and the post-relaxation crossover to long-time diffusion.

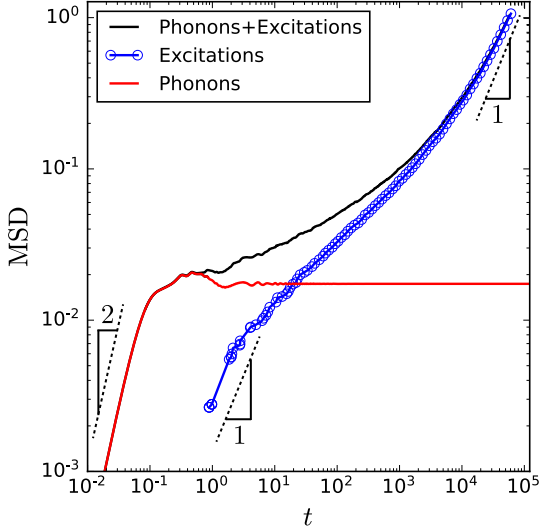

**Fig. 9.** The combined effect of phonon and excitation fluctuations in the MSD (black) along with the phonon (red) and the excitation part (blue), where the excitation part is obtained from Fig. 8(b) at $\beta J_\sigma = 3$ and $\kappa = 2/3$. The overall MSD shows the ballistic-subdiffusive-diffusive crossover behaviors typical of a supercooled liquid.

**The Role of Phonons.** In supercooled liquids, microscopic motion consists of vibrational fluctuations, i.e., phonons, around inherent states and excitation-mediated hopping between such states. The combination of excitations and phonons provides the complete picture of how a supercooled liquid relaxes at all timescales. To see this, suppose that a supercooled liquid behaves as a fluctuating linear elastic solid when it vibrates around an IS. This assumption leads to the following phononic MSD for a liquid with density $\rho$:

$$\mathrm{MSD}^{\mathrm{ph}}(t) = \frac{k_\mathrm{B} T}{4\pi\rho} \left[ \left( \frac{1}{c_\mathrm{s}^2} + \frac{1}{c_\mathrm{p}^2} \right) \ln \frac{L}{\sigma} \right.$$
$$+ \frac{1}{c_\mathrm{s}^2} \left[ \mathrm{Ci}\left(\frac{4\pi c_\mathrm{s} t}{\sigma}\right) - \mathrm{Ci}\left(\frac{4\pi c_\mathrm{s} t}{L}\right) \right]$$
$$\left. + \frac{1}{c_\mathrm{p}^2} \left[ \mathrm{Ci}\left(\frac{4\pi c_\mathrm{p} t}{\sigma}\right) - \mathrm{Ci}\left(\frac{4\pi c_\mathrm{p} t}{L}\right) \right] \right], \quad [30]$$

where $c_\mathrm{s}$ and $c_\mathrm{p}$ are the speed of shear and pressure waves, respectively, and $\mathrm{Ci}(s)$ is the cosine integral; see SM Sec. 6.1 for derivations. If phonon and excitation fluctuations are statistically independent, we may then write the overall MSD as

$$\mathrm{MSD}(t) = \mathrm{MSD}^{\mathrm{exc}}(t) + \mathrm{MSD}^{\mathrm{ph}}(t), \quad [31]$$

where the superscript 'exc' denotes excitation contributions from Eq. (29).

Figure 9 shows the overall MSD obtained from combining Eq. (30) with the kMC data from Fig. 8 at $\beta J_\sigma = 3$, choosing the frequency prefactor to be $\nu_0 = 0.1$ and $J_\sigma = 1.71$. At early times, we see that phonon contributions dominate those from excitations providing the initial ballistic regime of the overall MSD, given by Eq. (30) as $t \to 0$ yielding $\mathrm{MSD}^{\mathrm{ph}}(t) = \frac{2\pi k_\mathrm{B} T}{\rho\sigma^2} t^2$. The phonon part then plateaus to the equilibrium MSD of the solid, i.e., as $t \to \infty$, $\mathrm{MSD}^{\mathrm{ph}}(t) = \frac{k_\mathrm{B} T}{4\pi\rho} \left( \frac{1}{c_\mathrm{s}^2} + \frac{1}{c_\mathrm{p}^2} \right) \ln \frac{L}{\sigma}$. The presence of the logarithm scaling with system size $L$ indicates the effect of Mermin-Wagner fluctuations (37, 38). Meanwhile, the excitation contributions take over the phonon contributions at longer timescales, leading to an overall subdiffusion behavior at intermediate timescales, ultimately reaching the long-time diffusion behaviors around the relaxation times (Fig. 9). In effect, the overall MSD exhibits the ballistic-subdiffusive-diffusive crossover behaviors that are an important characteristic of the supercooled liquids (1). Note that as the temperatures are further lowered, the long-time asymptotic behavior of the phonons persists leading to a longer plateau-like behavior of the overall MSD, again consistent with observations (10, 42).

In addition to the MSD, the phonon contributions influence the relaxation functions. To that end, for independent phonon and excitation fluctuations, the overall self-intermediate scattering function can be written as the product of the phonon and excitation parts (SM Sec. 6.2):

$$F_\mathrm{s}(\mathbf{k}, t) = F_\mathrm{s}^{\mathrm{exc}}(\mathbf{k}, t) F_\mathrm{s}^{\mathrm{ph}}(\mathbf{k}, t), \quad [32]$$

$$F_\mathrm{s}^{\mathrm{ph}}(\mathbf{k}, t) = e^{-\frac{k^2}{4} \mathrm{MSD}^{\mathrm{ph}}(t)}. \quad [33]$$

The phonon part only depends on the corresponding MSD, since phonon fluctuations are Gaussian and all higher-order cumulants contributing to the self-intermediate scattering function vanish exactly. As time $t \to 0$ the phonon part $F_\mathrm{s}^{\mathrm{ph}}(\mathbf{k}, t) = \exp(-k^2 \frac{\pi k_\mathrm{B} T}{2\rho\sigma^2} t^2)$ exhibits a Gaussian decay, while its long-time limit reaches a plateau value that scales as a power law with linear system size $L$, i.e., $F_\mathrm{s}^{\mathrm{ph}}(\mathbf{k}, t) = \left(\frac{L}{\sigma}\right)^{-\eta}$ with $\eta = k^2 \frac{k_\mathrm{B} T}{16\pi\rho} \left( \frac{1}{c_\mathrm{s}^2} + \frac{1}{c_\mathrm{p}^2} \right)$. Similar decomposition can also be performed for the bond-order relaxation function (SM Sec. 6.3):

$$C_\mathrm{b}(t) = C_\mathrm{b}^{\mathrm{exc}}(t) C_\mathrm{b}^{\mathrm{ph}}(t), \quad [34]$$

$$C_\mathrm{b}^{\mathrm{ph}}(t) = e^{-18\Theta^{\mathrm{ph}}(t)}, \quad [35]$$

$$\Theta^{\mathrm{ph}}(t) = \frac{k_\mathrm{B} T}{8\pi\rho c_\mathrm{s}^2} \left[ \frac{\pi}{\sigma^2} + \frac{1 - \cos\left(\frac{4\pi c_\mathrm{s} t}{\sigma}\right)}{8\pi (c_\mathrm{s} t)^2} - \frac{\sin\left(\frac{4\pi c_\mathrm{s} t}{\sigma}\right)}{8\pi (c_\mathrm{s} t)^2} \right]. \quad [36]$$

Again, as $t \to 0$, the short time limit exhibits a Gaussian decay given by $C_\mathrm{b}^{\mathrm{ph}}(t) = \exp(-\frac{9\pi^3 k_\mathrm{B} T}{\rho\sigma^4} t^2)$, while the long-time limit reaches a constant plateau value of $\exp(-\frac{9\pi k_\mathrm{B} T}{4\rho c_\mathrm{s}^2 \sigma^2})$ free of finite-size effects.

Figure 10 shows the combined relaxation functions at low and high temperatures. At low temperatures, phonons provide the initial decay (green) towards a plateau value (red), accompanied by damped oscillations, before we observe the long-time decay mediated by the excitations. At high temperatures, the separation of timescales between the phonon and excitation part starts to vanish as the system completely relaxes because of excitations, and the two-step process becomes less apparent. The two-step process shown in Fig. 10 is characteristic of the relaxation behavior of supercooled liquids and is often observed in experiments and MD simulations (1, 10, 42).

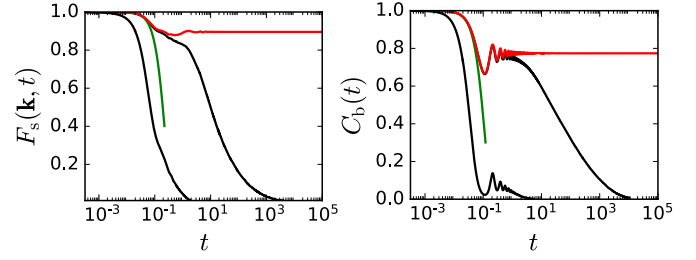

**Fig. 10.** The combined effect of phonon and excitation fluctuations in $F_\mathrm{s}(\mathbf{k}, t)$ (left) and $C_\mathrm{b}(t)$ (right), which are shown in black. Both relaxation functions exhibit the two-step decay commonly observed in supercooled liquids at low temperatures. The excitation part is obtained from kMC simulations at $\kappa = 3/8$ and $L = 15\ell_\mathrm{d}$. Left and right black curves correspond to $\beta J_\sigma = 0.44$ and $\beta J_\sigma = 3.86$, respectively. Also shown are green curves representing the short-time Gaussian decays of the relaxation functions for $\beta J_\sigma = 3.86$. The curves shown in red are the phononic contributions to the relaxation functions saturating to corresponding plateau values at long times for $\beta J_\sigma = 3.86$.

## Discussion

In summary, our work presents a theory for how dynamical facilitation emerges from excitations interacting elastically with mechanical stresses accumulated by previous excitations. With emergent facilitation, we reproduce key features of glassy dynamics including (1) the stretched exponential decay of relaxation functions, (2) the super-Arrhenius increase of relaxation timescales and their finite size effects in 2D, and (3) the emergence of subdiffusive behavior of MSD, all consistent with observations from molecular simulations and experiments. We also derive the phonon contribution to relaxation, which when added to the inherent state dynamics reproduces the two-step decay of relaxation functions.

Our model can be compared to other models incorporating facilitation. For instance, kinetically constrained models (KCMs) (13) represent a class of spin-lattice models and are used as coarse-grained models of supercooled liquids in the context of DF theory (3), where spins correspond to the excitations. Unlike standard lattice models, spin flips in KCMs only proceed around nearest-neighboring spins—a constraint that puts facilitation by hand. To obtain super-Arrhenius behavior, kinetics is further constrained; each spin must possess a unit vector, and spin flips only occur when the unit vector of neighboring spins has the same directionality (11, 43). Such a constraint leads to anisotropic facilitation despite the largely isotropic spreading of glassy dynamics observed in supercooled liquids (10). In contrast to KCMs, facilitation is an emergent property in our work, obtained solely from elastic interactions with no constraints, yet leading to isotropic mobility spreading (Fig. 4(b)) and super-Arrhenius behaviors at low temperatures. Our work may also be viewed, alternatively, as a refinement of the KCMs, wherein the stringent directionality constraints are removed and replaced with a self-consistent microscopic theory that leads to emergent facilitation and other features of glassy dynamics that KCMs may also predict.

Another class of models that incorporate facilitation are elastoplastic (EP) models. Developed initially for understanding plasticity in soft glassy materials (44), EP models have been recently applied to understand relaxation in supercooled liquids (45, 46). EP models use elastic fields induced by localized plastic events, which arrive irreversibly, to determine how stresses relax the system over time. To the best of our understanding, facilitation is again put in by hand and exists as yielding criteria, constraining dynamics in regions where stresses have yet to reach a critical threshold. EP models typically produce Arrhenius relaxation timescales and exponentially decaying relaxation functions (45, 46). Despite similar use of elastic fields, it is not yet clear how this class of models reproduces the stretched-exponential, super-Arrhenius, and sub-diffusive behaviors, which are defining features of equilibrium supercooled liquids.

Future work should focus on a quantitative comparison of the model to the relaxation data from experiments and molecular simulations. This depends on estimating the three model parameters energy barrier $J_\sigma$, elastic interaction strength $\kappa$ and lattice length scale $\ell_d$. There exist reasonable approximate estimates of these parameters for model polydisperse systems studied in MD simulations, with analytical formulas presented in this work and Ref. (5), upon which further refinements may be made. Since there exists a wealth of experimental data for three-dimensional (3D) systems, a quantitative comparison requires us to extend the model to 3D. Such a task, however, requires us first to understand the nature of localized excitations in 3D, thereby building a theory to predict $J_\sigma$ from the knowledge of structural and IS elastic properties. Nevertheless, the current work and the theory of localized excitations, applied in 2D and explored in Ref. (5), provide a blueprint for a microscopic-based theory of 3D glassy dynamics.

**ACKNOWLEDGMENTS.** M.R.H. and K.K.M. were supported by the Director, Office of Science, Office of Basic Energy Sciences, of the US Department of Energy under Contract No. DEAC02-05CH11231. This research used resources of the National Energy Research Scientific Computing Center (NERSC), a U.S. Department of Energy Office of Science User Facility located at Lawrence Berkeley National Laboratory, using NERSC award BES-ERCAP0023682. We acknowledge useful discussions with Sanggeun Song, Elizabeth Dresselhaus, Ashesh Ghosh, Amaresh Sahu, Dimitrios Fraggedakis, and Hyeongjoo Row.

# Supplemental Material
# Emergent Facilitation and Glassy Dynamics in Supercooled Liquids

Muhammad R. Hasyim[1,‡]   Kranthi K. Mandadapu[1,2,*]

[1]Department of Chemical & Biomolecular Engineering, University of California, Berkeley, CA, USA 94720

[2]Chemical Sciences Division, Lawrence Berkeley National Laboratory, Berkeley, CA, USA 94720

# Contents

[‡] muhammad_hasyim@berkeley.edu

[*] kranthi@berkeley.edu

# 1 Deriving the 2D Formula for Excitation Energy Cost (Eq. [10])

We begin with the general expression for the excitation energy cost in indicial notation,

$$J_\sigma = \frac{1}{2} \int d^2\mathbf{x} \, d_{ij}^\ddagger C_{ijkl}^{\text{IS}} d_{kl}^\ddagger, \tag{1.1}$$

where $d_{ij}^\ddagger = \epsilon_{ij}^\ddagger - \frac{1}{2}\epsilon_{kk}^\ddagger \delta_{ij}$ is the deviatoric part of $\epsilon_{ij}^\ddagger$, and $C_{ijkl}^{\text{IS}}$ is the IS elasticity tensor. For an isotropic medium, the IS elasticity tensor can be written as

$$C_{ijkl}^{\text{IS}} = \lambda^{\text{IS}} \delta_{ij} \delta_{kl} + G^{\text{IS}}(\delta_{ik}\delta_{jl} + \delta_{il}\delta_{jk}), \tag{1.2}$$

where $\lambda^{\text{IS}}$ is the first Lamé constant and $G^{\text{IS}}$ is the shear modulus. Multiplying Eq. (1.2) with $d_{kl}^\ddagger$ gives us the elastic deviatoric stress induced by the excitation

$$S_{ij}^\ddagger = C_{ijkl}^{\text{IS}} d_{kl}^\ddagger = \lambda^{\text{IS}} \delta_{ij} d_{kk}^\ddagger + G^{\text{IS}}(d_{ij}^\ddagger + d_{ji}^\ddagger) = 2G^{\text{IS}} d_{ij}^\ddagger. \tag{1.3}$$

Substituting Eq. (1.3) into Eq. (1.1) yields

$$J_\sigma = \int d^2\mathbf{x} \, G^{\text{IS}} d_{ij}^\ddagger d_{ij}^\ddagger. \tag{1.4}$$

In 2D, we can simplify Eq. (1.4) by writing the strain tensor in terms of its two unique deviatoric components. To achieve this, we note that given a symmetric tensor $A_{ij}$ and the following basis tensors

$$e_{ij}^{(1)} = \frac{1}{\sqrt{2}}\begin{bmatrix} -1 & 0 \\ 0 & -1 \end{bmatrix}, \quad e_{ij}^{(2)} = \frac{1}{\sqrt{2}}\begin{bmatrix} -1 & 0 \\ 0 & 1 \end{bmatrix}, \quad e_{ij}^{(3)} = \frac{1}{\sqrt{2}}\begin{bmatrix} 0 & 1 \\ 1 & 0 \end{bmatrix}, \tag{1.5}$$

we can write $A_{ij}$ as follows:

$$A_{ij} = \sum_{p=1}^{3} A_p e_{ij}^{(p)}, \tag{1.6}$$

$$A_1 = -\frac{A_{11} + A_{22}}{\sqrt{2}}, \quad A_2 = \frac{A_{22} - A_{11}}{\sqrt{2}}, \quad A_3 = \sqrt{2} A_{12}, \tag{1.7}$$

where $A_1$ is the hydrostatic component since $e_{ii}^{(p)} \neq 0$ when $p = 1$ and zero otherwise, leaving us with $A_2$ and $A_3$ as the deviatoric components of the tensor. Substituting Eq. (1.6) into Eq. (1.4), with $A_{ij} = d_{ij}^\ddagger$, we find that the hydrostatic part vanishes in $J_\sigma$ leaving us with all deviatoric parts of the strain energy:

$$J_\sigma = \int d^2\mathbf{x} \, G^{\text{IS}} \left[ \left(d_2^\ddagger\right)^2 + \left(d_3^\ddagger\right)^2 \right]. \tag{1.8}$$

Note that Eq. (1.8) is identical to Eq. [5] of the main text.

To obtain the final formula for $J_\sigma$ (Eq. [10] of the main text), we can use the force-dipole formalism [1] where an excitation corresponds to two pairs of opposing point forces [2], referred to as force dipoles, placed such that they induce localized pure shear. The details of using the force-dipole formalism to obtain the final formula for $J_\sigma$ are found in Ref. [2]. Here, we write the key result from Ref. [2] needed to derive $J_\sigma$, which is the elastic stress field of an excitation oriented at an angle $\psi$,

$$T_1^\ddagger(r, \theta; \psi) = \sqrt{2} f^\ddagger R_{\text{exc}} \frac{(\nu^{\text{IS}} + 1)}{\pi r^2} \sin(2\theta - 2\psi) \tag{1.9}$$

$$T_2^\ddagger(r,\theta;\psi) = \sqrt{2} f^\ddagger R_{\text{exc}} \frac{(\nu^{\text{IS}}+1)}{\pi r^2} \cos(4\theta - 2\psi), \tag{1.10}$$

$$T_3^\ddagger(r,\theta;\psi) = -\sqrt{2} f^\ddagger R_{\text{exc}} \frac{(\nu^{\text{IS}}+1)}{\pi r^2} \sin(4\theta - 2\psi), \tag{1.11}$$

where $f^\ddagger$ is the magnitude of the point forces, $(r,\theta)$ are the polar coordinates. Note that, in comparison to Eqs. (26)–(27) of Ref. [2], Eqs. (1.9)–(1.11) are adjusted by a factor of $\sqrt{2}$, to be consistent with the definitions in Eq. (1.7), and valid only for $r > 0$. Using Eqs. (1.9)–(1.11), we can write the components of the deviatoric stress tensor $S_{ij}^\ddagger = T_{ij}^\ddagger - \frac{1}{2}\delta_{ij} T_{kk}^\ddagger$ as follows:

$$S_1^\ddagger(r,\theta;\psi) = 0 \tag{1.12}$$

$$S_2^\ddagger(r,\theta;\psi) = \sqrt{2} f^\ddagger R_{\text{exc}} \frac{(\nu^{\text{IS}}+1)}{\pi r^2} \cos(4\theta - 2\psi), \tag{1.13}$$

$$S_3^\ddagger(r,\theta;\psi) = -\sqrt{2} f^\ddagger R_{\text{exc}} \frac{(\nu^{\text{IS}}+1)}{\pi r^2} \sin(4\theta - 2\psi). \tag{1.14}$$

The magnitude $f^\ddagger$ can be related to the eigenstrain $\epsilon_c$, which determines how much shear strain is experienced at the core of the excitation [2], and is written as

$$f^\ddagger = \frac{\sqrt{2}\pi R_{\text{exc}} G^{\text{IS}} \epsilon_c}{(1+\nu^{\text{IS}})} \tag{1.15}$$

where $\nu^{\text{IS}}$ is the IS Poisson's ratio. Using Eqs. (1.3) and (1.13)–(1.15), we can write the deviatoric components of $d_{ij}^\ddagger$ as

$$d_2^\ddagger(r,\theta;\psi) = \frac{\epsilon_c R_{\text{exc}}^2}{r^2} \cos(4\theta - 2\psi), \tag{1.16}$$

$$d_3^\ddagger(r,\theta;\psi) = -\frac{\epsilon_c R_{\text{exc}}^2}{r^2} \sin(4\theta - 2\psi). \tag{1.17}$$

Substituting Eqs. (1.16) and (1.17) and performing the integral with a short-distance cut off at $r = R_{\text{exc}}$, we obtain the final formula for the excitation energy cost

$$J_\sigma = G^{\text{IS}} \epsilon_c^2 \pi R_{\text{exc}}^2, \tag{1.18}$$

which is Eq. [10] of the main text.

## 2 Deriving the Interaction Energy (Eq. [12])

In this section, we derive the interaction term of the free-energy barrier $\Delta F_{21}^\ddagger$ to transition from a state with one excitation ($\alpha = 1$) to a state with two excitations ($\alpha' = 2$). Recall from the main text that the general expression for $\Delta F_{\alpha'\alpha}^\ddagger$ is given by

$$\Delta F_{\alpha'\alpha}^\ddagger = J_\sigma + \int d^2\mathbf{x}\, T_{ij}^\alpha d_{ij}^\ddagger, \tag{2.1}$$

where $d_{ij}^\ddagger$ is the deviatoric strain arising from a new bond-exchange event leading to the new state $\alpha'$. For the transition from one excitation to two excitations, the deviatoric strain field $d_{ij}^\ddagger$ corresponds to the elastic deviatoric strain field $d_{ij}^2(\mathbf{x} - \mathbf{q}^2; \psi^2)$ of an excitation, labeled as ②, that is positioned at $\mathbf{q}^2$ and oriented at an angle $\psi^2$ (Fig. 2c of the main text). The stress field $T_{ij}^\alpha$ corresponds to the elastic stress field $T_{ij}^1(\mathbf{x} - \mathbf{q}^1)$ of the excitation labeled as ① positioned at $\mathbf{q}^1$ and oriented at an angle $\psi^1$. Given this set of excitations, we can denote the interaction energy term in Eq. (2.1) as

$$v_{\text{int}}^{21} := \int d^2\mathbf{x}\, T_{ij}^1 d_{ij}^2. \tag{2.2}$$

We can write Eq. (2.2) in terms of displacement fields by applying the definition of a strain tensor $\epsilon_{ij} = \frac{1}{2}(u_{i,j} + u_{j,i})$, the divergence theorem, and integration by parts. The result is

$$v_{\text{int}}^{21} = -\frac{1}{2}\int d^2\mathbf{x}\, (T_{ij,i}^1 u_j^2 + T_{ij,j}^1 u_i^2) = -\frac{1}{2}\int d^2\mathbf{x}\, (f_j^1 u_j^2 + f_i^1 u_i^2) \tag{2.3}$$

where the last equation is obtained from the relation between stresses and external forces, $T_{ij,j}^1 = f_i^1$ where $f_i^1$ is the total force exerted by the force dipoles modeling excitation ①. In the force-dipole formalism, the expression for $f_i^1$ can be written as

$$f_i^1 = -P_{ij}^1 \delta_{,j}(\mathbf{x} - \mathbf{q}^1), \tag{2.4}$$

where $P_{ij}^1$ is the dipole moment tensor [1]. Substituting Eq. (2.4) into Eq. (2.3), we can perform the integration of the Dirac delta function to obtain

$$v_{\text{int}}^{21} = -P_2^1 d_2^2(\mathbf{q}^1 - \mathbf{q}^2; \psi^2) - P_3^1 d_3^2(\mathbf{q}^1 - \mathbf{q}^2; \psi^2). \tag{2.5}$$

The deviatoric strain fields $d_2^2$ and $d_3^2$ belong to excitation ② and are identical to the excitation strain fields $d_2^\ddagger$ and $d_3^\ddagger$ written in Eqs. (1.16) and (1.17), and thus

$$d_2^2(\mathbf{q}^1 - \mathbf{q}^2; \psi^2) = d_2^\ddagger(q^{21}, \theta^{21}; \psi^2), \quad d_3^2(\mathbf{q}^1 - \mathbf{q}^2; \psi^2) = d_3^\ddagger(q^{21}, \theta^{21}; \psi^2) \tag{2.6}$$

where $q^{21} = |\mathbf{q}^2 - \mathbf{q}^1|$ and $\theta^{21}$ is the polar angle of the vector $\mathbf{q}^2 - \mathbf{q}^1$. Equation (2.6) implies that Eq. (2.5) can be written as

$$v_{\text{int}}^{21} = -P_2^1 d_2^\ddagger(q^{21}, \theta^{21}; \psi^2) - P_3^1 d_3^\ddagger(q^{21}, \theta^{21}; \psi^2). \tag{2.7}$$

For the case of a 2D pure-shear excitation with radius $R_{\text{exc}}$, we can write the components of the dipole moment tensor as follows [2]:

$$P_1^1 = 0, \quad P_2^1 = -2\sqrt{2} f^1 R_{\text{exc}} \cos 2\psi^1, \quad P_3^1 = 2\sqrt{2} f^1 R_{\text{exc}} \sin 2\psi^1, \tag{2.8}$$

where $f^1$ and $\psi^1$ are the excitation force magnitude and orientation, respectively, of excitation ①. The force magnitude $f^1$ is set to $2f^\ddagger$—twice the magnitude at the transition state—since the excitation ① has crossed over the transition state. Using Eq. (1.15) and the fact that $J_\sigma = G^{\text{IS}}(\pi R_{\text{exc}}^2)\epsilon_c^2$, the components of the dipole moment tensor become

$$P_1^1 = 0, \quad P_2^1 = -\frac{8J_\sigma}{(1+\nu^{\text{IS}})\epsilon_c}\cos 2\psi^1, \quad P_3^1 = \frac{8J_\sigma}{(1+\nu^{\text{IS}})\epsilon_c}\sin 2\psi^1. \tag{2.9}$$

Substituting Eqs. (1.16), (1.17), (2.6) and (2.9) into Eq. (2.7), we obtain

$$v_{\text{int}}^{21} = \frac{2J_\sigma(2R_{\text{exc}})^2}{(1+\nu^{\text{IS}})(q^{21})^2}\left[\cos 2\psi^1 \cos(4\theta^{21} - 2\psi^2) + \sin 2\psi^1 \sin(4\theta^{21} - 2\psi^2)\right] \tag{2.10}$$

$$= \frac{\kappa J_\sigma}{(\tilde{q}^{21})^2}\cos\left(2\psi^2 + 2\psi^{21} - 4\theta^{21}\right) \tag{2.11}$$

where $\tilde{q}^{21} = q^{21}/2R_{\text{exc}}$, and $\kappa = \frac{2}{1+\nu^{\text{IS}}}$. Since the interaction diverges at $\tilde{q}^{21} = 0$, the interaction energy requires a cut-off lengthscale, which may be set to $2R_{\text{exc}}$. Equation (2.11) is Eq. [12] in the main text.

## 3 Minima of the Interaction Energy (Eq. [13])

To minimize Eq. (2.11), we can extremize over the pairwise distance and orientation angles. Following the main text, we assume that the position and orientation of the first excitation are fixed to $\mathbf{q}^1$ and $\psi^1$, respectively. The optimal orientation angle of the second excitation $\psi^{2*}$ can be obtained by minimizing the cosine term, i.e., when the following condition holds:

$$\cos\left(2\psi^{2*} + 2\psi^1 - 4\theta^{21}\right) = -1. \tag{3.1}$$

Equation (3.1) provides a range of solutions, as one can shift the polar angle $\theta^{21}$ of the pairwise distance vector and still obtain a valid solution. Letting $\theta^{21}$ as a free parameter, we can write the solution for Eq. (3.1) as follows:

$$\psi^{2*} = 2\theta^{21} + \pi\left(n + \frac{1}{2}\right) - \psi^1, \tag{3.2}$$

where $n$ is an integer. At this minimum, the interaction potential is an attractive well, i.e., $v^{21}_{\text{int}} = -\kappa J_\sigma/\left(\tilde{q}^{21}\right)^2$, with a hard-wall cutoff at $\tilde{q}^{21} = 1$. This implies that the minimizing pairwise distance is

$$\tilde{q}^{21*} = 1. \tag{3.3}$$

Equations (3.2) and (3.3) constitute Eq. [13] of the main text.

## 4 The Kinetic Monte Carlo (kMC) Algorithm

The kinetic Monte Carlo (kMC) algorithm [3] is a method to simulate continuous-time Markov processes with discrete state space. A master equation describes the dynamics of such processes:

$$\frac{\mathrm{d}p_\alpha(t)}{\mathrm{d}t} = \sum_{\alpha'}\left[w_{\alpha\alpha'}p_{\alpha'}(t) - w_{\alpha'\alpha}p_\alpha(t)\right] \tag{4.1}$$

where $p_\alpha(t)$ is the probability of being in state $\alpha$ at time $t$ and $w_{\alpha'\alpha}$ is the transition rate. Examples of Markov processes described by Eq. (4.1) include $d$-dimensional random walks on a lattice and the spin dynamics of the Ising model. Given the transition rates, the kMC algorithm simulates the Markov process by choosing a new state to jump towards from an initial state, repeating this process, and thus generating a trajectory of states as well as their arrival times $\{\alpha_k, t_k\}_{k=0}^K$, where $K$ is the number of jumps set by the algorithm.

A key element to the kMC algorithm is the probability function $p(\alpha', \tau; \alpha)$ to observe a new state $\alpha'$ given that the system remains at state $\alpha$ for some waiting time $\tau$. For the discrete-space continuous-time Markov processes, this probability function can be written as [3]

$$p(\alpha', \tau; \alpha) = w_{\alpha'\alpha}e^{-\tau W_\alpha}, \tag{4.2}$$

where $W_\alpha = \sum_{\alpha''} w_{\alpha''\alpha}$ is the total transition rate. Summing Eq. (4.2) over all possible states gives us the probability distribution of waiting times,

$$\sum_{\alpha'} p(\alpha', \tau; \alpha) = p(\tau; \alpha) = W_\alpha e^{-\tau W_\alpha}, \tag{4.3}$$

**Algorithm 1:** The Kinetic Monte Carlo (kMC) Algorithm

**Data:** Initial conditions $\alpha_0$ and time $t_0$, total number of jumps $K$. Initial transition rates $w_{\alpha_1 \alpha_0}$ and total rate $W_{\alpha_0}$

1 **for** $k = 0, \ldots, K$ **do**
2 $\quad$ Sample $\alpha_{k+1} \sim p(\alpha_{k+1}; \alpha_k) = \frac{w_{\alpha_{k+1}\alpha_k}}{W_{\alpha_k}}$
3 $\quad$ Sample $\tau_{k+1} \sim p(\tau_{k+1}; \alpha_k) = W_{\alpha_k} e^{-\tau_{k+1} W_{\alpha_k}}$
4 $\quad$ Store $\alpha_{k+1}, \tau_{k+1}$
5 $\quad$ Recompute $w_{\alpha_{k+1}\alpha_k}$ and $W_{\alpha_k}$ (if necessary)

Figure S.1: Pseudo-code for the kMC algorithm.

which is an exponential distribution. By definition, the conditional probability $p(\alpha' \mid \tau; \alpha)$ to observe the new state $\alpha'$ given that the waiting time is $\tau$ is given by

$$p(\alpha' \mid \tau; \alpha) := \frac{p(\alpha', \tau; \alpha)}{p(\tau; \alpha)} = \frac{w_{\alpha'}}{\sum_{\alpha''} w_{\alpha''\alpha}} \, . \tag{4.4}$$

Since the RHS of Eq. (4.4) is independent of $\tau$, the probability of choosing a new state $\alpha'$ is statistically independent of the waiting times — a vital feature of the Markov property. By exploiting this statistical independence, we can write a simple algorithm described in Fig. S.1 for simulating discrete-space continuous-time Markov processes.

## 4.1 Implementation of the Markov State Model

The theory presented in the main text describes relaxation dynamics by the arrival of excitations in continuous space, which is incompatible with the discrete nature of the kMC algorithm. Before we consider the discrete version of the theory, we briefly summarize the transition rates in continuous space. To that end, in continuous space, the transition rate for creating a new excitation, labeled $\mu$, at position $\mathbf{q}^\mu$ and orientation angle $\psi^\mu$, given $n$-many excitations is

$$w_{\alpha'\alpha} = \nu_0 e^{-\beta \Delta F_{\alpha'\alpha}^\ddagger}, \quad \Delta F_{\alpha'\alpha}^\ddagger = J_\sigma + \sum_{\mu' \neq \mu} v_{\text{int}}^{\mu\mu'}, \quad \text{where } \alpha \to \alpha' \text{ increases } n \to n+1, \tag{4.5}$$

$$v_{\text{int}}^{\mu\mu'} = \frac{\kappa J_\sigma}{(\tilde{q}^{\mu\mu'})^2} \cos\left(2\psi^\mu + 2\psi^{\mu'} - 4\theta^{\mu\mu'}\right) . \tag{4.6}$$

As described in the main text, the free-energy barrier for deleting an excitation is the same as the barrier for creating that same excitation. In other words, the transition rate for deleting an excitation labeled $\mu$ is

$$w_{\alpha'\alpha} = \nu_0 e^{-\beta \Delta F_{\alpha'\alpha}^\ddagger}, \quad \Delta F_{\alpha'\alpha}^\ddagger = J_\sigma + \sum_{\mu' \neq \mu} v_{\text{int}}^{\mu\mu'}, \quad \text{where } \alpha \to \alpha' \text{ decreases } n \to n-1. \tag{4.7}$$

where $v_{\text{int}}^{\mu\mu'}$ is the same as Eq. (4.6). The continuous nature of space and orientation angle changes the form of the total transition rate, which now involves an integral

$$\frac{W_\alpha}{k_{\text{exc}}} = \sum_\mu e^{-\beta \sum_{\mu' \neq \mu} v_{\text{int}}^{\mu\mu'}} + \int_\Omega \frac{d^2\mathbf{q}^\mu}{(2R_{\text{exc}})^2} \int_0^{2\pi} d\psi^\mu e^{-\beta \sum_{\mu' \neq \mu} v_{\text{int}}^{\mu\mu'}} \tag{4.8}$$

where $k_{\text{exc}} = \nu_0 e^{-\beta J_\sigma}$, $\Omega$ defines a region where the new excitation does not overlap with an existing excitation by a diameter $2R_{\text{exc}}$.

To construct the discrete version of the model, we imagine triangulating the 2D space, with the lattice spacing being $\ell_\text{d}$, so that excitations only occupy $N_\ell$-many lattice sites. Let us denote the set of lattice sites as $\{\mathbf{x}_\mu\}_{\mu=1}^{N_\ell}$. The resulting triangular lattice transforms the integral in Eq. (4.8) into a sum, which we can use to deduce the transition rates appropriate for the kMC algorithm. To write the discrete version of Eq. (4.8), we define $n_\mu$ as the occupation variable for the $\mu$-th lattice site, which is zero if no excitation exists at position $\mathbf{x}_\mu$ and one otherwise, and $\mathcal{E} = \{\mu \in \mathbb{N} : n_\mu = 1\}$ is the set of indices where excitations occupy the corresponding sites. With these definitions, the discrete total rate can be written as

$$\frac{\hat{W}_\alpha}{k_{\text{exc}}} = \sum_{\mu \notin \mathcal{E}} A e^{-\beta \sum_{\mu' \neq \mu} v_{\text{int}}^{\mu\mu'}} + \sum_{\mu \in \mathcal{E}} \int_0^{2\pi} d\psi^\mu A e^{-\beta \sum_{\mu' \neq \mu} v_{\text{int}}^{\mu\mu'}}. \tag{4.9}$$

Here, $A = 1$ if the site is occupied, and $A = \frac{\sqrt{3}}{2} \frac{\ell_\text{d}^2}{(2R_{\text{exc}})^2}$ if the site is unoccupied, with the latter also equal to the area of a hexagon centered at a site. If we choose a lattice spacing, where $\ell_\text{d} \sim O(2R_{\text{exc}})$, we can write the interaction energy using a new dimensionless pairwise distance $\hat{q}^{\mu\mu'} = q^{\mu\mu'}/\ell_\text{d}$ as

$$v_{\text{int}}^{\mu\mu'} = \frac{\hat{\kappa} J_\sigma}{(\hat{q}^{\mu\mu'})^2} \cos\left(2\psi^\mu + 2\psi^{\mu'} - 4\theta^{\mu\mu'}\right) \tag{4.10}$$

where $\hat{\kappa} = \kappa(\frac{2R_{\text{exc}}}{\ell_\text{d}})^2$. From Eq. (4.9), we see that the discrete transition rate, denoted as $\hat{w}_{\alpha'\alpha}$, can be written as

$$\hat{w}_{\alpha'\alpha} = A\nu_0 e^{-\beta \Delta F_{\alpha'\alpha}^\ddagger}, \quad \Delta F_{\alpha'\alpha}^\ddagger = J_\sigma + \sum_{\mu' \neq \mu} v_{\text{int}}^{\mu\mu'}. \tag{4.11}$$

In this work, we shall treat the interaction strength $\hat{\kappa}$ as an adjustable parameter.

## 4.2 Applying the kMC Algorithm

With Eqs. (4.10) and (4.11) at hand, we can implement the kMC algorithm as written in Fig. S.1. The first step of the kMC algorithm involves sampling of a new state given the current state $\alpha$ of the system. Given state $\alpha_k$ at the $k$-th step of the algorithm, this step decides whether the algorithm should delete or create an excitation. This decision can be made by assigning a probability $P^\mu(\alpha)$ to every $\mu$-th lattice site that tells us the likelihood of flipping the occupation variable $n_\mu$, given that the system starts from state $\alpha$. If state $\alpha$ has $n$-many excitations, the probability function can be obtained from Eq. (4.11) by dividing the rate expression with the appropriate normalizing constant $Z_\alpha$. Additionally, the probability $P^\mu(\alpha)$ must consider the total rate of creating excitations at any angle, which we can achieve via an integral over the orientation angle. The result of following these steps is as follows:

$$P^\mu(\alpha) = \begin{cases} \frac{1}{Z_\alpha} \int_0^{2\pi} d\psi^\mu \ A e^{-\beta \sum_{\mu' \neq \mu} v_{\text{int}}^{\mu\mu'}} & \text{if } n_\mu = 0 \\ \frac{1}{Z_\alpha} A e^{-\beta \sum_{\mu' \neq \mu} v_{\text{int}}^{\mu\mu'}} & \text{otherwise.} \end{cases} \tag{4.12}$$

$$Z_\alpha = \sum_{\mu'' \notin \mathcal{E}} A e^{-\beta \sum_{\mu' \neq \mu''} v_{\text{int}}^{\mu''\mu'}} + \sum_{\mu'' \in \mathcal{E}} \int_0^{2\pi} d\psi^{\mu''} A e^{-\beta \sum_{\mu' \neq \mu''} v_{\text{int}}^{\mu''\mu'}} \tag{4.13}$$

where $Z_\alpha$ is the normalizing constant. Once the algorithm decides the lattice site, whose label is denoted by $\mu$, an additional step is needed if a new excitation is created, i.e., sampling an orientation

**Algorithm 2:** The kMC Algorithm for the Markov State Model

**Data:** Initial conditions $\alpha_0$ and time $t_0$, total number of jumps $K$, initial probability $p(\mu; \alpha_0)$ and total rate $\hat{W}_{\alpha_0}$.

1. **for** $k = 0, \ldots, K$ **do**
2.     Sample $\mu^* \sim P^\mu(\alpha_k)$ via `numpy.random.choice`
3.     **if** $n_{\mu^*} = 0$ **then**
4.         Compute $\rho(\psi^\mu) = \frac{e^{-\beta V^\mu(\psi)}}{\int_0^{2\pi} d\psi' \ e^{-\beta V^\mu(\psi')}}$
5.         Sample $\psi^* \sim \rho(\psi^\mu)$ using ITS.
6.         Insert excitation at $\mathbf{q}^{\mu^*}$ with orientation $\psi^\mu = \psi^*$
7.     **else**
8.         Delete excitation at $\mathbf{q}^{\mu^*}$
9.     Compute $\tau_{k+1} = \frac{1}{\hat{W}_{\alpha_k}} \ln(1/(1-u))$ where $u \sim \text{Unif}[0, 1]$
10.    Update state $\alpha_k \to \alpha_{k+1}$ and time $t_{k+1} = t_k + \tau_{k+1}$
11.    Update the lattice-site probabilities $P^\mu(\alpha_k) \to P^\mu(\alpha_{k+1})$ and the total rate $\hat{W}_{\alpha_k} \to \hat{W}_{\alpha_{k+1}}$ (Fig. S.3).

Figure S.2: Pythonic pseudo-code for the kMC algorithm as applied to the Markov state model of this work.

angle $\psi^\mu$ for the new excitation. We denote the probability distribution for the orientation angle as $\rho(\psi^\mu)$, which we can obtain from normalizing the rate expression in Eq. (4.11) at fixed $\mu$ as

$$\rho(\psi^\mu) = \frac{e^{-\beta \sum_{\mu' \neq \mu} v_{\text{int}}^{\mu\mu'}}}{\int_0^{2\pi} d\psi^{\mu''} \ e^{-\beta \sum_{\mu' \neq \mu''} v_{\text{int}}^{\mu''\mu'}}} \ . \tag{4.14}$$

Equations (4.12) and (4.14) constitute a complete set of probability functions that we can use to perform the sampling step in the kMC algorithm.

Sampling from Eq. (4.14) and Eq. (4.12) can be done through several algorithms, the most common of which is the inverse transform sampling (ITS). Given a 1D continuous probability distribution $\rho(x)$, we can sample a random variable $X$ using the ITS in four steps:

1. Generate a random number $u \sim \text{Unif}[0, 1]$, where $\text{Unif}[0, 1]$ is the uniform distribution.
2. Compute the cumulative distribution function (CDF), $F_X(x) := \int_{x_{\min}}^x dx' \ \rho(x')$.
3. Find the inverse of the desired CDF, i.e., $F_X^{-1}(u)$.
4. Compute $X = F_X^{-1}(u)$.

In this work, we use the ITS to sample from Eq. (4.14) and the exponential distribution for waiting times. The inverse of the CDF of Eq. (4.14) is obtained numerically, where a trapezoidal rule is used with a set of $N_\psi$-many gridpoints to approximate the following CDF:

$$F_\psi(x) = \frac{\int_0^x d\psi^\mu \ e^{-\beta \sum_{\mu' \neq \mu} v_{\text{int}}^{\mu\mu'}}}{\int_0^{2\pi} d\psi^{\mu''} \ e^{-\beta \sum_{\mu' \neq \mu''} v_{\text{int}}^{\mu''\mu'}}} \ , \tag{4.15}$$

**Algorithm 3:** Updating the lattice-site probabilities and total rate

`Data:` Lattice site $\mu^*$ where an excitation has been created/deleted. New orientation angle $\psi^*$ if creating an excitation. Gridpoints for orientation angle $\{\psi_i\}_{i=1}^{N_\psi}$. Current total interaction energy per site $V_k^\mu(\psi)$.

1 `if inserting excitation then`
2     # Updating sites that will be occupied
3     `if` $n_\mu > 0$ `for any` $\mu$ `then`
4        # For old occupied sites. Vectorize over $\mu$
5        Update $V_{k+1}^\mu(\psi^\mu) = V_k^\mu(\psi^\mu) + v_{\text{int}}^{\mu\mu^*}$ for $\mu \in \mathcal{E}$ and $\mu \neq \mu^*$
6        # For the newly occupied site $\mu^*$. Vectorize over $\mu'$.
7        Compute $V_{k+1}^{\mu^*}(\psi^{\mu^*}) = \sum_{\substack{\mu' \in \mathcal{E}, \\ \mu' \neq \mu^*}} v_{\text{int}}^{\mu^*\mu'}$
8     `else`
9        Compute $V_{k+1}^{\mu^*}(\psi^{\mu^*}) = 0$
10    Compute unnormalized probability $z_{k+1}^\mu = e^{-\beta V_{k+1}^\mu(\psi^\mu)}$ for $\mu \in \mathcal{E}$ and $\mu = \mu^*$
11    Set $n_{k+1}^{\mu^*} = 1$ and $\psi_{k+1}^{\mu^*} = \psi^*$
12    # Updating sites that remain unoccupied
13    # For all empty sites. Vectorize over $\mu$ and $\psi$
14    Update $V_{k+1}^\mu(\psi) = V_k^\mu(\psi) + v_{\text{int}}^{\mu\mu^*}$ for $\mu \notin E$ and $\psi \in \{\psi_i\}_{i=1}^{N_\psi}$
15    Compute unnormalized probability $z_{k+1}^\mu = \int_0^{2\pi} d\psi \, \frac{\sqrt{3}}{2} \left(\frac{\ell_d}{2R_{\text{exc}}}\right)^2 e^{-\beta V_{k+1}^\mu(\psi)}$.
16 `else`
17    # Updating sites that will be unoccupied
18    # For all old empty sites. Vectorize over $\mu$ and $\psi$
19    Update $V_{k+1}^\mu(\psi) = V_k^\mu(\psi) - v_{\text{int}}^{\mu\mu^*}$ for $\mu \in \mathcal{E}$, $\mu \neq \mu^*$, and $\psi \in \{\psi_i\}_{i=1}^{N_\psi}$
20    Set $n_\mu = 0$ and remove $\psi^{\mu^*}$
21    `if` $n_\mu > 0$ `for any` $\mu$ `then`
22        Update $V_{k+1}^{\mu^*}(\psi^{\mu^*}) = \sum_{\substack{\mu' \in \mathcal{E}, \\ \mu' \neq \mu^*}} v_{\text{int}}^{\mu^*\mu'}$
23    `else`
24        Update $V_{k+1}^{\mu^*}(\psi^{\mu^*}) = 0$
25    Compute unnormalized probability $z_{k+1}^\mu = \int_0^{2\pi} d\psi \, \frac{\sqrt{3}}{2} \left(\frac{\ell_d}{2R_{\text{exc}}}\right)^2 e^{-\beta V_{k+1}^\mu(\psi)}$.
26    # Updating sites that remain occupied
27    `if` $n_\mu > 1$ `then`
28        # For old occupied sites $\mu$. Vectorize over $\mu$
29        Update $V_{k+1}^\mu(\psi^\mu) = V_k^\mu(\psi^\mu) - v_{\text{int}}^{\mu\mu^*}$ for $\mu \notin \mathcal{E}$
30        Compute unnormalized probability $z_{k+1}^\mu = e^{-\beta V_{k+1}^\mu(\psi^\mu)}$
31    # Compute total transition rate and normalized probability
32    Compute $Z_{\alpha_{k+1}} = \sum_{\mu=1}^{N_\ell} z_{k+1}^\mu$, $\hat{W}_{\alpha_{k+1}} = \nu_0 e^{-\beta J_\sigma} Z_{\alpha_{k+1}}$, and $P^\mu(\alpha_{k+1}) = \frac{z_{k+1}^\mu}{Z_{\alpha_{k+1}}}$

Figure S.3: Pythonic pseudo-code for updating the transition probabilities of the Markov state model. This algorithm enters the last step of Algorithm 1 in Fig. S.2.

which we then invert by swapping the input and output values. For the waiting times, we can find the inverse of the CDF analytically, and it is written as

$$F_\tau^{-1}(u) = \frac{1}{W_{\alpha_k}} \ln(1/(1-u)). \tag{4.16}$$

Note that the ITS can also be applied to a discrete probability distribution, e.g., Eq. (4.12), via interpolation. For efficiency, however, we sample from Eq. (4.12) using the built-in `numpy.random.choice` function from Python.

The last step of the kMC algorithm is to update the lattice-site probabilities $P^\mu(\alpha_k) \to P^\mu(\alpha_{k+1})$ and the total rate $\hat{W}_{\alpha_k} \to \hat{W}_{\alpha_{k+1}}$. This step is the most expensive portion of the kMC algorithm due to the long-range nature of interactions. Given $n$-many excitations, updating $P^\mu(\alpha_k)$ for empty sites requires us to recompute the interaction energy, which takes $n$-many operations, for every gridpoint in $\psi$. Updating $P^\mu(\alpha_k)$ for occupied sites also requires us to recompute the interaction energy, which takes $n$-many operations, but for a single orientation $\psi^\mu$. In total, the complexity of this step is of order $O(n(N_\ell - n)N_\psi + n^2) \sim O(n^2)$.

The complexity can be reduced by implementing incremental updates to the probabilities and total rates as the kMC algorithm marches in time. This is possible by storing an array of total interaction energies per site, denoted as $V^\mu$. For every site $\mu$, we need to store $(N_\psi+1)$-many values of $V^\mu$, corresponding to $N_\psi$-many for the gridpoints $\{\psi_i\}_{i=1}^{N_\psi}$, which are used in the integration over $\psi$, and one more for the orientation angle $\psi^\mu$ if an excitation occupies the $\mu$-th lattice site. Given $V^\mu$, the incremental updates can be done following Fig. S.3, where the complexity at every $k$-th step of the kMC algorithm is now reduced to $O(2n + N_\psi(N_\ell - n)) \sim O(n)$ and $O(nN_\psi + 2n) \sim O(n)$ for creating and deleting an excitation, respectively. Combined with `numpy` built-in vectorization for their arrays, the Markov state model can be efficiently simulated in Python for a wide range of temperatures.

## 4.3 Parameters for Simulations

Table S.1: Model parameters that are kept constant in all kMC simulations. Values are reported in Lennard-Jones units, as described for a model polydisperse system in Ref. [2] named Poly-(12,0).

| $\ell_\mathrm{d}$ | $R_\mathrm{exc}$ | $u^\ddagger$ | $\sigma$ | $G^\mathrm{IS}$ | $\nu^\mathrm{IS}$ | $\rho$ |
|---|---|---|---|---|---|---|
| $\frac{4}{3}\sigma$ | $\frac{\sqrt{3}}{2}\sigma$ | $0.116\sigma$ | 1.1 | 11 | 0.35 | 1.01 |

All kMC simulations are done with parameters either kept fixed or varied systematically. We keep several parameters constant throughout the kMC simulation, and they are listed in Table S.1—all figures reported in the main text utilize these constant parameters. Varied parameters include the linear system size $L$, interaction strengths $\kappa$, and inverse temperature $\beta$. For every figure in the main text, they are varied as follows:

- Fig. 4: $L = 100\ell_\mathrm{d}$, $\kappa = 1/2$, and $\beta J_\sigma = 2$ and $\beta J_\sigma = 5$ for Fig. 4(a) and (b), respectively.

- Fig. 5: $L = 15\ell_\mathrm{d}$, $\kappa = 1/2$, and the inverse temperature is varied within the range $\beta J_\sigma \in [0, 3.86]$.

- Fig. 6: $L = 15\ell_\mathrm{d}$, interaction strength is varied at values $\kappa \in \{3/8, 1/2, 2/3\}$, and inverse temperature is varied within the range $\beta J_\sigma \in [0, 4]$.

- Fig. 7: The system size is varied at values $\frac{L}{\ell_\mathrm{d}} \in \{30, 60, 100, 200\}$, $\kappa = 1/2$, and the inverse temperature is varied within the range $\beta J_\sigma \in [0, 3]$.

- Fig. 8: $L = 15\ell_d$, $\kappa = 2/3$, and the inverse temperature is varied at values $\beta J_\sigma \in \{2.5, 2.0, 3.0\}$.
- Fig. 9: $L = 15\ell_d$, $\kappa = 2/3$, and $\beta J_\sigma = 3.0$.
- Fig. 10: $L = 15\ell_d$, $\kappa = 3/8$, and inverse temperature is varied at values $\beta J_\sigma \in \{0.44, 3.86\}$.

# 5 Deriving the Displacement Field of an Excitation

In this section, we derive the elastic Green's function $\mathbf{u}^{\text{exc}}$ for an excitation that is positioned at $\mathbf{q}$ and oriented at an angle $\psi$. From the force-dipole formalism of linear elasticity theory [1,2], we can write the elastic Green's function as follows:

$$u_i^{\text{exc}} = -G_{ij,k}(\mathbf{x} - \mathbf{q})P_{jk}(\psi) \tag{5.1}$$

where $G_{ij}(\mathbf{x})$ is the elastic Green's function for a unit point force at the origin,

$$G_{ij} = \frac{1}{4\pi G^{\text{IS}}} \left[ \frac{1+\nu^{\text{IS}}}{2} \frac{x_i x_j}{|\mathbf{x}|^2} - \left(\frac{3-\nu^{\text{IS}}}{2}\right) \ln|\mathbf{x}|\delta_{ij} \right], \tag{5.2}$$

and $P_{jk}$ is the dipole moment tensor for an excitation with force magnitude $2f^\ddagger$:

$$P_{jk} = 4f^\ddagger R_{\text{exc}} \tilde{P}_{jk}, \quad \tilde{P}_{jk} = \begin{bmatrix} \cos 2\psi & \sin 2\psi \\ \sin 2\psi & -\cos 2\psi \end{bmatrix}. \tag{5.3}$$

Deriving $\mathbf{u}^{\text{exc}}$ requires computing the partial derivative of Eq. (5.2), which is

$$G_{ij,k} = \frac{1}{4\pi G^{\text{IS}}} \left[ \frac{1+\nu^{\text{IS}}}{2} \frac{\partial}{\partial x_k} \left(\frac{x_i x_j}{|\mathbf{x}|^2}\right) - \left(\frac{3-\nu^{\text{IS}}}{2}\right) \frac{\partial}{\partial x_k} \ln|\mathbf{x}|\delta_{ij} \right] \tag{5.4}$$

$$= \frac{1}{4\pi G^{\text{IS}}|\mathbf{x}|} \tilde{G}_{ijk} \tag{5.5}$$

where $\tilde{G}_{ijk}$ is given by

$$\tilde{G}_{ijk} = \left[ \frac{1+\nu^{\text{IS}}}{2}(\hat{x}_j \delta_{ik} + \hat{x}_i \delta_{jk} - 2\hat{x}_i \hat{x}_j \hat{x}_k) - \left(\frac{3-\nu^{\text{IS}}}{2}\right) \delta_{ij}\hat{x}_k \right]. \tag{5.6}$$

Substituting Eqs. (5.5) and (5.6) into Eq. (5.1) we obtain

$$u_i^{\text{exc}}(r,\theta) = -\frac{f^\ddagger R_{\text{exc}}}{\pi G^{\text{IS}} r} \tilde{G}_{ijk}(\theta) \tilde{P}_{jk}(\psi). \tag{5.7}$$

where $r = |\mathbf{x} - \mathbf{q}|$ and $\theta$ is the polar angle of $\mathbf{x} - \mathbf{q}$. We can derive a more compact form by turning $\tilde{P}_{jk}(\psi)$ into a 'director' vector $\mathbf{n} = [\cos 2\psi, \sin 2\psi]$. This is possible by writing $\tilde{P}_{jk}$ as follows

$$\tilde{P}_{jk} = \cos 2\psi \, s_{jk}^{(1)} + \sin 2\psi \, s_{jk}^{(2)}, \quad \text{with} \tag{5.8}$$

$$\mathbf{s}^{(1)} = \begin{bmatrix} 1 & 0 \\ 0 & -1 \end{bmatrix}, \quad \mathbf{s}^{(2)} = \begin{bmatrix} 0 & 1 \\ 1 & 0 \end{bmatrix}. \tag{5.9}$$

Substituting Eqs. (5.8) and (5.9) into Eq. (5.7), we obtain in vector notation

$$\begin{bmatrix} u_x^{\text{exc}} \\ u_y^{\text{exc}} \end{bmatrix} = -\frac{f^\ddagger R_{\text{exc}}}{\pi G^{\text{IS}} r} \begin{bmatrix} \tilde{G}_{1j,k} s_{jk}^{(1)} & \tilde{G}_{1j,k} s_{jk}^{(2)} \\ \tilde{G}_{2j,k} s_{jk}^{(1)} & \tilde{G}_{2j,k} s_{jk}^{(2)} \end{bmatrix} \begin{bmatrix} \cos 2\psi \\ \sin 2\psi \end{bmatrix} \tag{5.10}$$

$$\implies \mathbf{u}^{\text{exc}} = -\frac{f^{\ddagger}R_{\text{exc}}}{\pi G^{\text{IS}}r}\tilde{\mathbf{G}}\cdot\mathbf{n}, \qquad (5.11)$$

where $\tilde{G}$ can be written more compactly as follows:

$$\tilde{\mathbf{G}} = \begin{bmatrix} \tilde{G}_{1j,k}s_{jk}^{(1)} & \tilde{G}_{1j,k}s_{jk}^{(2)} \\ \tilde{G}_{2j,k}s_{jk}^{(1)} & \tilde{G}_{2j,k}s_{jk}^{(2)} \end{bmatrix} \qquad (5.12)$$

$$= \begin{bmatrix} \cos\theta\left(\nu^{\text{IS}} - 1 - (1+\nu^{\text{IS}})\cos 2\theta\right) & \sin\theta\left(-2 - (1+\nu^{\text{IS}})\cos 2\theta\right) \\ \sin\theta\left(1 - \nu^{\text{IS}} - (1+\nu^{\text{IS}})\cos 2\theta\right) & \cos\theta\left(-2 + (1+\nu^{\text{IS}})\cos 2\theta\right) \end{bmatrix}. \qquad (5.13)$$

Carrying out the matrix multiplication $\tilde{\mathbf{G}}\cdot\mathbf{n}$ in Eq. (5.11), we obtain

$$\mathbf{u}^{\text{exc}}(r,\theta;\psi) = \frac{f^{\ddagger}R_{\text{exc}}}{\pi G^{\text{IS}}r}\begin{bmatrix} \frac{3-\nu^{\text{IS}}}{2}\cos(\theta-2\psi) + \frac{1+\nu^{\text{IS}}}{2}\cos(3\theta-2\psi) \\ -\frac{3-\nu^{\text{IS}}}{2}\sin(\theta-2\psi) + \frac{1+\nu^{\text{IS}}}{2}\sin(3\theta-2\psi) \end{bmatrix}. \qquad (5.14)$$

From the displacement field, we can then compute the rotational strain field $\theta^{\text{exc}} = \frac{1}{2}\varepsilon_{ij}u_{i,j}^{\text{exc}}$. The result is

$$\theta^{\text{exc}}(r,\theta;\psi) = -\frac{2f^{\ddagger}R_{\text{exc}}}{\pi G^{\text{IS}}r^2}\sin(2\theta-2\psi). \qquad (5.15)$$

Equation (5.15) is useful for computing the bond-order autocorrelation function in the system.

The parameter $f^{\ddagger}$ has a formula, written in Eq. (1.15), used throughout our computational studies. However, we may also use an alternative formula based on setting the displacement magnitude to a particular value. Note that the magnitude of Eq. (5.14) can be written as

$$u(r,\theta) := |\mathbf{u}^{\text{exc}}(r,\theta)| = \frac{f^{\ddagger}R_{\text{exc}}}{\sqrt{2}G^{\text{IS}}\pi r}\sqrt{(5 + (\nu^{\text{IS}}-2)\nu^{\text{IS}} + (3-\nu^{\text{IS}})(1+\nu^{\text{IS}})\cos(4\theta-4\psi)} \qquad (5.16)$$

which depends on both orientation and angle. If we look at $\theta = \psi$ and $r = R_{\text{exc}}$, we have

$$u(r = R_{\text{exc}}, \theta = \psi) = \frac{2f^{\ddagger}}{G^{\text{IS}}\pi}. \qquad (5.17)$$

Setting the force magnitude $f^{\ddagger}$ so that the displacement magnitude is equal to $2u^{\ddagger}$ gives us the following:

$$f^{\ddagger} = \pi u^{\ddagger}G^{\text{IS}} \qquad (5.18)$$

and so upon substituting Eq. (5.18) into Eq. (5.14) we obtain

$$\mathbf{u}^{\text{exc}}(r,\theta;\psi) = \frac{u^{\ddagger}R_{\text{exc}}}{r}\begin{bmatrix} \frac{3-\nu^{\text{IS}}}{2}\cos(\theta-2\psi) + \frac{1+\nu^{\text{IS}}}{2}\cos(3\theta-2\psi) \\ -\frac{3-\nu^{\text{IS}}}{2}\sin(\theta-2\psi) + \frac{1+\nu^{\text{IS}}}{2}\sin(3\theta-2\psi) \end{bmatrix}. \qquad (5.19)$$

Equation (5.19) may be useful as it allows us to interpret displacement magnitudes in terms of the distance required to displace bonds leading up to the transition state in a bond-exchange event.

# 6 Deriving the Phonon Contributions

In this section, we derive the phonon contribution to the relaxation dynamics. Assuming that the system obeys linear elasticity, the displacement fluctuations about an inherent state $\mathbf{u}(\mathbf{x}, t)$ resulting from thermal energy obey the elastic wave equation:

$$\rho \frac{\partial^2 u_i}{\partial t^2} = T_{ij,j} = (\lambda^{\text{IS}} + G^{\text{IS}})u_{j,ji} + G^{\text{IS}} u_{i,jj}, \tag{6.1}$$

with $\rho$ being the density. Let $u_i^{\mathbf{k}}(t) = \int d^2\mathbf{x}\, e^{-i\mathbf{k}\cdot\mathbf{x}} u(\mathbf{x}, t)$ be the Fourier transform of the displacement field with wavenumber $\mathbf{k}$. The Fourier transform of Eq. (6.1) can then be written as

$$\rho \ddot{u}_i^{\mathbf{k}}(t) = -(\lambda^{\text{IS}} + G^{\text{IS}})k_i k_j u_j^{\mathbf{k}}(t) - G^{\text{IS}} k^2 u_i^{\mathbf{k}}(t). \tag{6.2}$$

The system undergoes random motion through initial velocities sampled from the Maxwell-Boltzmann distribution.

To obtain the phonon contribution to relaxation, we first need the analytical solution to Eq. (6.2). Let $U_i^{\mathbf{k}}(s) = \int_0^\infty dt\, e^{-st} u_i^{\mathbf{k}}(t)$ be the Laplace transform of $u_i^{\mathbf{k}}(t)$. Now, taking the Laplace transform of Eq. (6.2), we find that

$$\rho\left[s^2 U_i^{\mathbf{k}}(s) - s u_i^{\mathbf{k}}(0) - \dot{u}_i^{\mathbf{k}}(0)\right] = -(\lambda^{\text{IS}} + G^{\text{IS}})k_i k_j U_j^{\mathbf{k}}(s) - G^{\text{IS}} k^2 U_i^{\mathbf{k}}(s). \tag{6.3}$$

Applying the initial conditions $u_i^{\mathbf{k}}(0) = 0$ and $\dot{u}_i^{\mathbf{k}}(0) = v_i^{\mathbf{k}}$, where $v_i^{\mathbf{k}}$ is the initial Boltzmann distributed velocities, we obtain

$$\left[\left(\rho s^2 + G^{\text{IS}} k^2\right)\delta_{ij} + (\lambda^{\text{IS}} + G^{\text{IS}})k_i k_j\right] U_j^{\mathbf{k}}(s) = \rho v_i^{\mathbf{k}} \tag{6.4}$$

The terms in the parentheses are $2\times 2$ matrices that can be inverted analytically and written as

$$G_{ij} = \frac{1}{\rho s^2 + G^{\text{IS}} k^2}\left[\delta_{ij} - \frac{(\lambda^{\text{IS}} + G^{\text{IS}})k_i k_j}{\rho s^2 + (\lambda^{\text{IS}} + 2G^{\text{IS}})k^2}\right], \tag{6.5}$$

leading to the displacement field in Fourier-Laplace space as $U_i^{\mathbf{k}}(s) = \rho G_{ij}(\mathbf{k}, s) v_j^{\mathbf{k}}$. Applying the inverse Laplace transform then gives us the following solution for $u_i^{\mathbf{k}}(t)$:

$$u_i^{\mathbf{k}}(t) = \frac{v_i^{\mathbf{k}} - \hat{k}_i \hat{k}_j v_j^{\mathbf{k}}}{c_{\text{s}} k} \sin(c_{\text{s}} k t) + \frac{\hat{k}_i \hat{k}_j v_j^{\mathbf{k}}}{c_{\text{p}} k} \sin(c_{\text{p}} k t), \tag{6.6}$$

where $c_{\text{s}} = \sqrt{\frac{G^{\text{IS}}}{\rho}}$ and $c_{\text{p}} = \sqrt{\frac{\lambda^{\text{IS}} + 2G^{\text{IS}}}{\rho}}$ are the speed of shear and pressure waves, respectively.

## 6.1 Mean Squared Displacement

From Eq. (6.6), we can derive several relaxation properties, starting with the mean-squared displacement. The MSD can be computed from the covariance of the displacement field as

$$\langle u_i(\mathbf{x}, t) u_j(\mathbf{x}, t)\rangle^{\text{ph}} = \int \frac{d^2\mathbf{k}}{(2\pi)^2} \int \frac{d^2\mathbf{k}'}{(2\pi)^2} \left\langle u_i^{\mathbf{k}}(t) u_j^{\mathbf{k}'}(t)\right\rangle^{\text{ph}} e^{i(\mathbf{k}+\mathbf{k}')\cdot\mathbf{x}}, \tag{6.7}$$

where $\langle\ldots\rangle^{\text{ph}}$ is an ensemble average over the Maxwell-Boltzmann distribution for the velocities in $\mathbf{k}$-space. Substituting Eq. (6.6) into Eq. (6.7) and performing the ensemble average, we obtain

$$\langle u_i(\mathbf{x}, t) u_j(\mathbf{x}, t)\rangle^{\text{ph}} = \frac{k_{\text{B}} T}{4\pi\rho} \int_{2\pi/L}^{2\pi/\sigma} \frac{dk}{k}\left[\frac{\delta_{ij} - \hat{k}_i \hat{k}_j}{c_{\text{s}}^2}(1 - \cos 2c_{\text{s}} k t) + \frac{\hat{k}_i \hat{k}_j}{c_{\text{p}}^2}(1 - \cos 2c_{\text{p}} k t)\right], \tag{6.8}$$

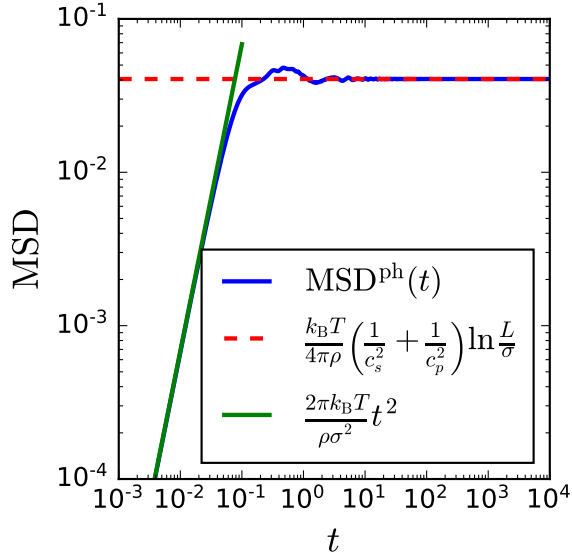

Figure S.4: The phonon MSD (blue line), its short-time ballistic limit (green line), and the plateau value (red dashed line). Parameters: $L = 15$, $k_\mathrm{B}T = 1.33$, $\sigma = 1.1$, $\rho = 1.01$, $G^\mathrm{IS} = 11.0$, and $\nu^\mathrm{IS} = 0.36$.

where $L$ is the linear system size and $\sigma$ is a characteristic particle diameter. The integration of the cosine term is related to a special function known as the cosine integral $\mathrm{Ci}(s) = -\int_s^\infty \mathrm{d}t\, \frac{\cos t}{t}$. Taking the trace of Eq. (6.8), we obtain the phonon part of the MSD:

$$\mathrm{MSD}^\mathrm{ph}(t) = \frac{k_\mathrm{B}T}{4\pi\rho}\left[\left(\frac{1}{c_\mathrm{s}^2} + \frac{1}{c_\mathrm{p}^2}\right)\ln\left(\frac{L}{\sigma}\right) + \frac{1}{c_\mathrm{s}^2}\left[\mathrm{Ci}\left(\frac{4\pi c_\mathrm{s} t}{\sigma}\right) - \mathrm{Ci}\left(\frac{4\pi c_\mathrm{s} t}{L}\right)\right]\right.$$
$$\left.+\frac{1}{c_\mathrm{p}^2}\left[\mathrm{Ci}\left(\frac{4\pi c_\mathrm{p} t}{\sigma}\right) - \mathrm{Ci}\left(\frac{4\pi c_\mathrm{p} t}{L}\right)\right]\right]. \tag{6.9}$$

As seen in Fig. S.4, the MSD shows different regimes. At short times, the MSD enters the ballistic regime, and the $\sim t^2$ scaling can be derived from Eq. (6.9) from a Taylor series expansion at $t = 0$:

$$\mathrm{MSD}^\mathrm{ph}(t) = \frac{2\pi k_\mathrm{B}T}{\rho\sigma^2}t^2 - \frac{(c_\mathrm{s}^2 + c_\mathrm{p}^2)2\pi^3 k_\mathrm{B}T}{3\rho\sigma^2}t^4 + O(t^6) \tag{6.10}$$

Next, the MSD undergoes damped oscillations before decaying towards a plateau value. Since cosine integrals decay to zero as $t \to \infty$, we can write the plateau value as the following long-time limit:

$$\lim_{t\to\infty} \mathrm{MSD}^\mathrm{ph}(t) = \frac{k_\mathrm{B}T}{4\pi\rho}\left(\frac{1}{c_\mathrm{s}^2} + \frac{1}{c_\mathrm{p}^2}\right)\ln\left(\frac{L}{\sigma}\right), \tag{6.11}$$

which is consistent with the equilibrium MSD obtained for a 2D continuum fluctuating solid. The logarithmic system-size dependence shifts the MSD to larger values as $L$ increases (Fig. S.5(right)) and is a unique feature for 2D solids. Temperature also plays a role in determining the plateau value (Fig. S.5(left)), with increasing temperature leading to larger MSD values.

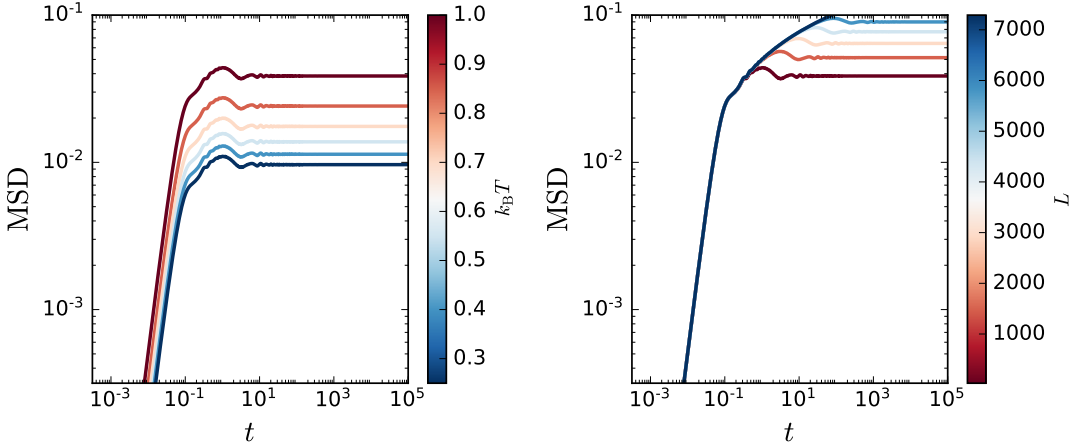

Figure S.5: The phonon MSD when temperature (left) and system size (right) are varied. Parameters: $\sigma = 1.1$, $\rho = 1.01$, $G^{\mathrm{IS}} = 11.0$, and $\nu^{\mathrm{IS}} = 0.36$.

## 6.2 Self Intermediate Scattering Function

Since the displacement is linear in the initial velocity, which is Gaussian distributed, the statistics of displacement fields at any time $t$ is also Gaussian. This helps us compute the phonon contribution for the self-intermediate scattering function

$$F_{\mathrm{s}}^{\mathrm{ph}}(\mathbf{k}, t) = \frac{1}{L^2} \int \mathrm{d}^2\mathbf{x} \, \left\langle e^{i\mathbf{k}\cdot\mathbf{u}(\mathbf{x},t)} \right\rangle^{\mathrm{ph}} . \tag{6.12}$$

where the Gaussianity implies the following:

$$F_{\mathrm{s}}^{\mathrm{ph}}(\mathbf{k}, t) = e^{-\frac{k^2}{4}\mathrm{MSD}^{\mathrm{ph}}(t)} . \tag{6.13}$$

As seen in Fig. S.6, $F_{\mathrm{s}}^{\mathrm{ph}}(\mathbf{k}, t)$ also different regimes. At short times, we have the decay due to the ballistic regime as follows:

$$F_{\mathrm{s}}^{\mathrm{ph}}(\mathbf{k}, t) = e^{-k^2 \frac{\pi k_{\mathrm{B}} T}{2\rho\sigma^2} t^2} + O(t^4) . \tag{6.14}$$

This decay switches towards a damped oscillatory behavior that vanishes towards a plateau value, which from Eq. (6.11) is given by

$$\lim_{t\to\infty} F_{\mathrm{s}}^{\mathrm{ph}}(\mathbf{k}, t) = \left(\frac{L}{\sigma}\right)^{-\eta} , \quad \eta = k^2 \frac{k_{\mathrm{B}} T}{16\pi\rho} \left(\frac{1}{c_{\mathrm{s}}^2} + \frac{1}{c_{\mathrm{s}}^2}\right) . \tag{6.15}$$

This result is consistent with the power-law finite-size scaling obtained from an equilibrium 2D solid [4]. Similar to the MSD, increasing $L$ results in a smaller plateau value (Fig. S.7(right)), and thus the system can relax faster by increasing system size. Increasing temperature also leads to the same effect (Fig. S.7(left)). The finite-size effects observed in the MSD and $F_{\mathrm{s}}(\mathbf{k}, t)$ are unique to 2D solids; the long-range phonon fluctuations responsible for this effect are the Mermin-Wagner fluctuations [5].

## 6.3 Bond-Order Autocorrelation Function

A derivation similar to $F_{\mathrm{s}}^{\mathrm{ph}}(\mathbf{k}, t)$ can be done for the bond-order correlation function

$$C_{\mathrm{b}}(t) = \frac{1}{L^2} \int \mathrm{d}^2\mathbf{x} \, \left\langle e^{i6\theta(\mathbf{x},t)} \right\rangle^{\mathrm{ph}} , \tag{6.16}$$

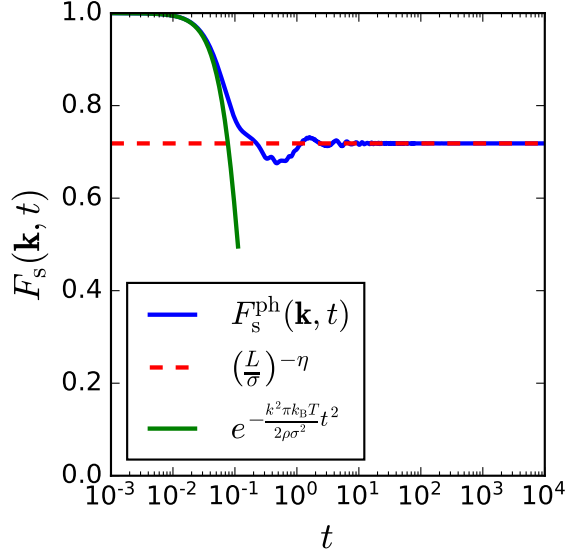

Figure S.6: The phonon $F_s(\mathbf{k},t)$ (blue line), its short-time ballistic limit (green line), and the plateau value (red dashed line). Parameters: $L = 15$, $k_B T = 1.33$, $\sigma = 1.1$, $\rho = 1.01$, $G^{\text{IS}} = 11.0$, and $\nu^{\text{IS}} = 0.36$.

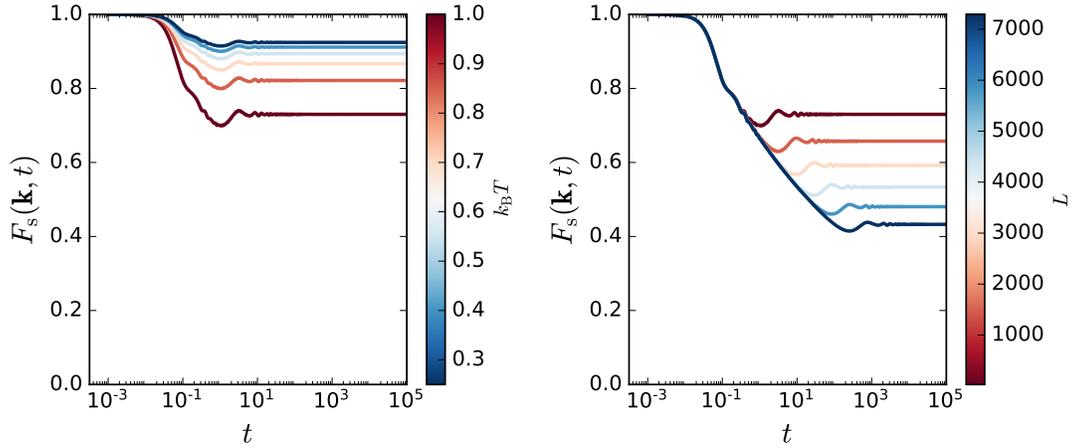

Figure S.7: The phonon $F_s(\mathbf{k},t)$ when temperature (left) and system size (right) are varied. Parameters: $\sigma = 1.1$, $\rho = 1.01$, $G^{\text{IS}} = 11.0$, and $\nu^{\text{IS}} = 0.36$.

where $\theta(\mathbf{x},t) = \frac{1}{2}\left[\frac{\partial u_x(\mathbf{x},t)}{\partial y} - \frac{\partial u_y(\mathbf{x},t)}{\partial x}\right] = \frac{1}{2}\varepsilon_{ij}u_{i,j}$ is the rotational strain, with $\varepsilon_{ij}$ being the 2D Levi-Civita tensor. Since $\theta(\mathbf{x},t)$ is also linear in displacement, its statistics are also Gaussian. Consequently, the bond-order correlation function becomes

$$C_\mathrm{b}(t) = e^{-18\Theta^\mathrm{ph}(t)}, \quad \Theta^\mathrm{ph}(t) := \langle\theta(\mathbf{x},t)\theta(\mathbf{x},t)\rangle^\mathrm{ph} = \frac{1}{4}\langle\varepsilon_{ij}u_{i,j}\varepsilon_{kl}u_{k,l}\rangle^\mathrm{ph}, \tag{6.17}$$

where $\Theta^\mathrm{ph}(t)$ is the covariance of rotational strain. In Fourier space, it can be computed as follows:

$$\Theta^\mathrm{ph}(t) = -\frac{1}{4}\int\frac{\mathrm{d}^2\mathbf{k}\,\mathrm{d}^2\mathbf{k}'}{(2\pi)^4}\varepsilon_{ij}k_j\varepsilon_{kl}k'_l\left\langle u_i^\mathbf{k}(t)u_k^{\mathbf{k}'}(t)\right\rangle^\mathrm{ph} e^{(\mathbf{k}+\mathbf{k}')\cdot\mathbf{x}}. \tag{6.18}$$

Note that the solution for the displacement is already found in Eq. (6.6). Once we substitute Eq. (6.6) into Eq. (6.18) and perform the ensemble average, we obtain the following:

$$\Theta^\mathrm{ph}(t) = \frac{kT}{8\pi\rho}\int_{2\pi/L}^{2\pi/\sigma}\mathrm{d}k\,k\,\varepsilon_{ij}k_j\varepsilon_{kl}k'_l\left[\frac{\delta_{ik}-\hat{k}_i\hat{k}_k}{c_s^2}\sin^2 c_s kt + \frac{\hat{k}_i\hat{k}_k}{c_p^2}\sin^2 c_p kt\right], \tag{6.19}$$

which can be further reduced to

$$\Theta^\mathrm{ph}(t) = \frac{k_\mathrm{B}T}{8\rho c_s^2}\left[\frac{1-\cos(4\pi c_s t/\sigma)}{8\pi(c_s t)^2} + \frac{\pi}{\sigma^2} - \frac{\sin(4\pi c_s t/\sigma)}{2\sigma(c_s t)}\right]. \tag{6.20}$$

Figure S.8 shows different regimes of $C_\mathrm{b}^\mathrm{ph}(t)$. By expanding Eq. (6.20) via a Taylor series at $t=0$, we recover the initial decay of the bond-order correlation function resulting from ballistic motion:

$$C_\mathrm{b}(t) = e^{-\frac{9\pi^3 k_\mathrm{B}T}{\rho\sigma^4}t^2} + O(t^4). \tag{6.21}$$

This behavior, again as in the self-intermediate scattering function, then switches to an oscillatory decay to a plateau value given by the long-time limit:

$$\lim_{t\to\infty} C_\mathrm{b}(t) = e^{-\frac{9\pi k_\mathrm{B}T}{4\rho c_s^2 \sigma^2}}, \tag{6.22}$$

which is independent of system size (Fig. S.9(right)). The absence of finite-size effects corroborates the general practice that $C_\mathrm{b}(t)$ is appropriate for studying relaxation in 2D supercooled liquids, free of the artifacts from the Mermin-Wagner fluctuations. Increasing temperature, however, still leads to a decrease in the plateau value, as shown in Fig. S.9(left).

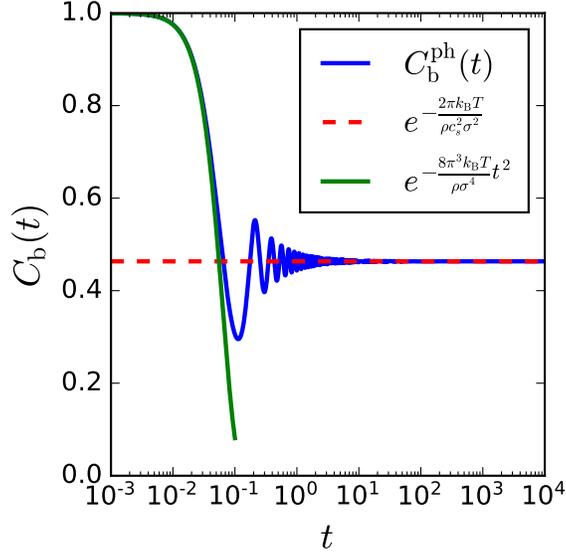

Figure S.8: The phonon $C_b(t)$ (blue line), its short-time ballistic limit (green line), and the plateau value (red dashed line). Parameters: $L = 15$, $k_B T = 1.33$, $\sigma = 1.1$, $\rho = 1.01$, $G^{IS} = 11.0$, and $\nu^{IS} = 0.36$.

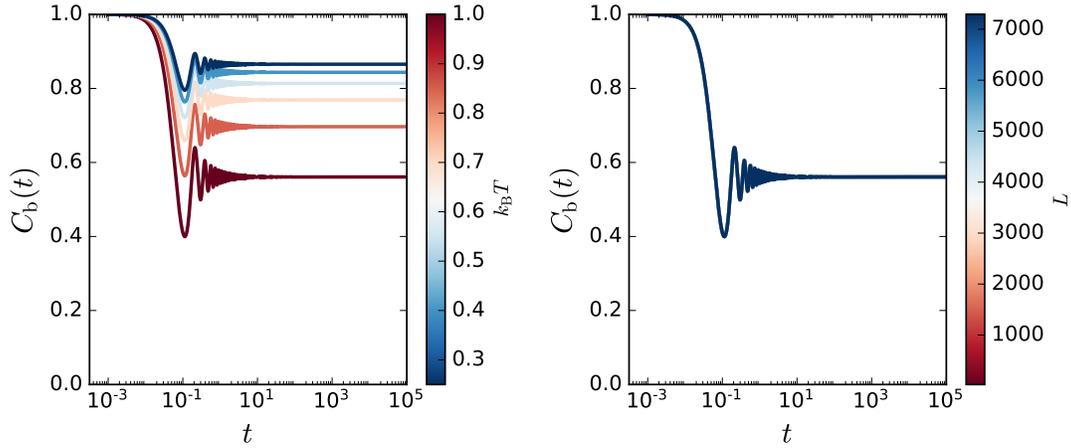

Figure S.9: The phonon $C_b(t)$ when temperature (left) and system size (right) are varied. Parameters: $\sigma = 1.1$, $\rho = 1.01$, $G^{IS} = 11.0$, and $\nu^{IS} = 0.36$.

# 7 Supplemental Figures

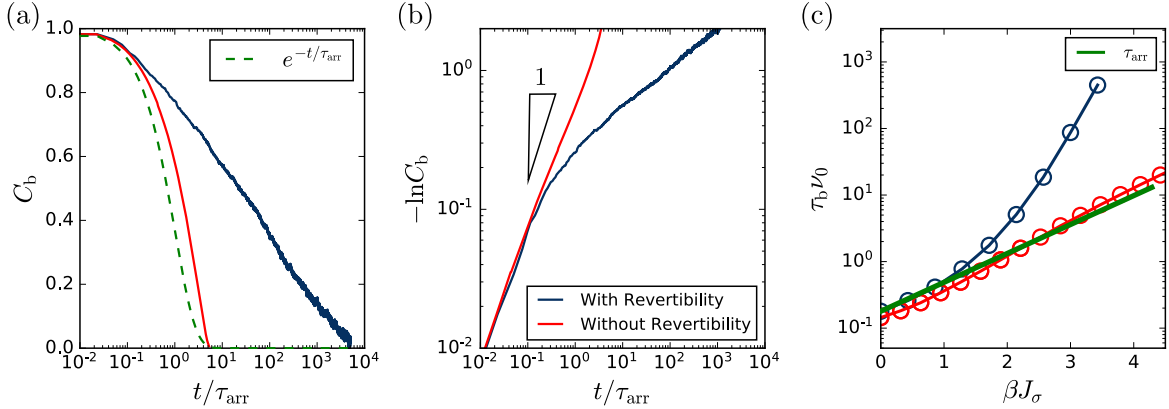

Figure S.10: Comparison between relaxation with (dark blue curve) and without (red curve) revertibility of excitations. The bond-order autocorrelation function as it decays over time at $\beta J_\sigma \approx 3$ and $\kappa = 2/3$ shown in linear-log scale (a) and in log-log scale (b). Note that for the case without revertibility, the decay of $C_b(t)$ is nearly exponential, just like the Poissonian case. The relaxation timescales obtained from $C_b(t = \tau_b) = 0.1$ are shown in (c), where Arrhenius behavior with energy barrier $J_\sigma$ is observed for all temperatures with no revertibility of excitations.

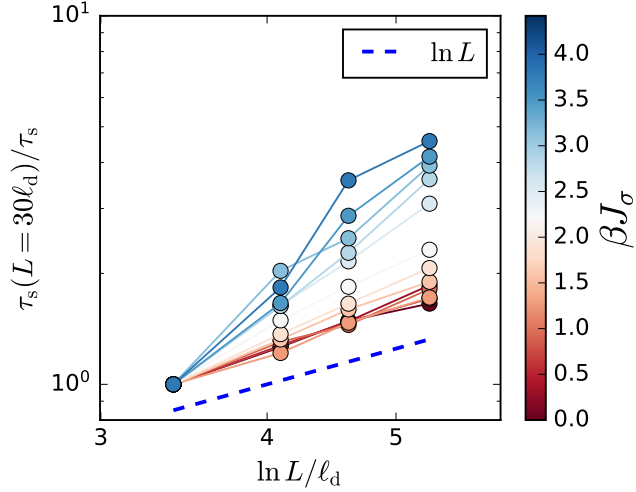

Figure S.11: Relaxation time data obtained from $F_s(\mathbf{k}, t)$ where we choose $|\mathbf{k}| = 2\pi/\sigma$. We see at high temperatures ($\beta J_\sigma \leq 2$) that the data can be collapsed to reveal a scaling $1/\tau_s \sim \ln L$, consistent with the analysis of the main text.

# 8 Supplemental Movies

Here, we provide links and descriptions for movies referred to in the main text.

- Movie S.1: https://youtube.com/shorts/niX0UI0RPX8?feature=share Time evolution of the persistence variable $p_\mu$, which tracks when an excitation first arrives at lattice site $\mu$, from a kMC simulation at high temperatures ($\beta J_\sigma = 1$). The parameters used in the simulation are the same as those in Fig. 4(a) of the main text. The video shows how excitations arrive randomly in space and time, resembling the behavior of a Poisson point process. Note that the Poissonian behavior continues until excitations almost completely occupy the system.

- Movie S.2: https://youtube.com/shorts/X7KmXTzqOtc?feature=share Time evolution of the persistence variable $p_\mu$, which tracks when an excitation first arrives at lattice site $\mu$, from a kMC simulation at low temperatures ($\beta J_\sigma = 5$). The parameters used in the simulation are the same as those in Fig. 4(b) of the main text. The video shows how dynamically active regions start from the initial set of excitations and coarsen over time. The coarsening process is a crucial signature of dynamical facilitation, as seen in MD simulations of model polydisperse glass formers [6].

- Movie S.3: https://youtube.com/shorts/WBypZiorO-4?feature=share Time evolution of the displacement field magnitude $|\mathbf{u}(\mathbf{x}, t)|$ from a kMC simulation at high temperatures ($\beta J_\sigma = 1$). The parameters used in the simulation and color coding in the video are the same as those in Fig. 4(c) of the main text.

- Movie S.4: https://youtube.com/shorts/EpCPDxhNyes?feature=share Time evolution of the magnitude of the displacement field $|\mathbf{u}(\mathbf{x}, t)|$ from a kMC simulation at low temperatures ($\beta J_\sigma = 5$). The parameters used in the simulation and color coding in the video are the same as those in Fig. 4(d) of the main text. At these low temperatures, displacement fields build up more slowly as excitations are constantly being reformed within the dynamically active regions, whose locations strongly correlate with regions with the highest magnitude of displacement.

- Movie S.5: https://youtu.be/IzTul8P6-PE Time evolution of the excitations and persistence variable $p_\mu$, colored orange and red, respectively, within a single active region. The kMC simulation is obtained at the same temperature and parameters as Movie S.1 and S.2. The video shows how the slow coarsening of the active region results from repeated reorganization of the regions that have already experienced bond-exchange events.

- Movie S.6: https://youtu.be/bNv1AnZuG1A Time evolution of the probe used in the kMC simulations at $\beta J_\sigma = 3.0$, showing caged displacements with intermittent hops. Note that the simulation parameters are the same as Fig. 8 of the main text.